\documentclass[prb,aps,twocolumn,superscriptaddress,floats,showpacs,amsmath,amssymb]{revtex4}

\usepackage{graphicx}
\usepackage{units}
\usepackage{bm}
\usepackage{epstopdf}
\def \be{\begin{equation}}
\def \ee{\end{equation}}

\def\u{\uparrow}
\def\d{\downarrow}
\def\Tr{\mbox{Tr}}
\def \h{"hat"}

\newcommand{\saclay}{%
           Nanoelectronics group, Service de Physique de l'Etat Condens{\'e}, 
           CEA Saclay F-91191 Gif-sur-Yvette Cedex, France}
\newcommand{\grenoble}{SPSMS, UMR-E 9001, CEA-INAC/ UJF-Grenoble 1,17 rue des martyrs, 38054 Grenoble cedex 9, France}
\newcommand{\wuerzburg}{%
           Institute for Theoretical Physics and Astrophysics,
University of W\"urzburg, D-97074 W\"urzburg, Germany}
\begin{document}
\title{Multiscale approach to spin transport in magnetic multilayers.}

\author{Simone Borlenghi}
\affiliation{\saclay}
\author{Valentin Rychkov}
\affiliation{\saclay}\affiliation{\wuerzburg}
\author{Cyril Petitjean}
\affiliation{\grenoble}
\author{Xavier Waintal}
\affiliation{\saclay}\affiliation{\grenoble}

\date{\today}

\begin{abstract}
This article discusses two dual approaches to spin transport in magnetic multilayers: a direct, purely quantum, approach based on a Tight-Binding model (TB) and a semiclassical approach (Continuous Random Matrix Theory, CRMT). The combination of both approaches provides a systematic way to perform multi-scales simulations of systems that contain relevant physics at scales larger (spin accumulation, spin diffusion...) and smaller (specular reflexions, tunneling...) than the elastic mean free paths of the layers. We show explicitly that CRMT and TB give consistent results in their common domain of applicability. 
\end{abstract}


\maketitle
\section{Introduction}

Many interesting physical effects can be observed in non collinear magnetic systems where the various magnetizations are not aligned or even not in the same plane. Indeed, in addition to the Giant Magneto Resistance (GMR~\cite{Baibich:1988,Binasch:1989}) and Tunneling  Magneto Resistance (TMR~\cite{Julliere:1975,Moodera:1995,Miyazaki:1995})  effects observed in the collinear configuration, the non conservation of  spin current gives rise to an influence of electronic transport on the magnetization dynamics which itself can lead to various phenomena such as magnetization reversal~\cite{Katine:2000,Fert:2004,Devolder:2007,Strachan:2008} or dc driven radio frequency oscillators~\cite{Kiselev:2003,Pufall:2006,Boone:2009,Houssameddine:2009, Villard:2010}. The potential for applications opened by these effects (magnetic memories, reprogrammable logic, tunable rf sources...) has been recognized very early and lead to an important expansion of the field~\cite{Chappert:2007}.

Consequently, the theory of non collinear magnetic systems has received a lot of attention.
Many different approaches have been developed, ranging from purely quantum~\cite{Braun:2004,Duine:2007} to semi-classical Boltzmann equation~\cite{Stiles:2002a,Xiao:2004,Xiao:2007}, (generalized) circuit theory~\cite{Brataas:2000, Brataas:2001, Bauer:2003a, Bauer:2003b, Tserkovnyak:2005,Brataas:2006, Manchon:2006} or Random Matrix Theory (RMT)~\cite{Waintal:2000, Rychkov:2009} or full ab-initio (density functional theory) calculations~\cite{Haney:2007a,Haney:2007b,Waldron:2007, Xu:2008}. A good understanding has now been reached of the respective domains of applicability and links between the various approaches~\cite{Brataas:2006,Rychkov:2009}. On the other hand it is becoming increasingly clear that the magnetization dynamics of real devices cannot be properly captured by simple (analytically tractable) models and that numerical simulations that treat the magnetic and transport degrees of freedom on an equal footing need to be developed. Several steps have already been taken in that direction~\cite{Fidler:2000, Lee:2004, Xiao:2005, Fischbacher:2007, Berkov:2008} and should eventually lead to simulations of good predictive capabilities.  Let us mention, as example,  the recent experiments on spin torque induced ferromagnetic resonance in the nonlinear regime~\cite{Chen:2009} which has been successfully compared to  coupled simulations
at the macrospin level for the magnetic part and one-dimensional semi-classical level for spin transport.  Other geometries with stronger spin texture (domain wall, vortex, strong oersted field...) will require a full three dimensional treatment however.

This paper is devoted to a discussion of a multiscale approach to the spin transport theory of magnetic multilayers: for some system, such as magnetic tunnel junctions, semi-classical approaches fail (they are intrinsically Ohmic and cannot capture the exponentially small transparency due to the presence of the insulating oxides) and one is therefore tempted to use  fully
quantum mechanical simulations. Those simulations are extremely computationally costly however, and cannot be performed on full size realistic devices. The multi-scale approach taken in this paper consists of describing the intrinsically quantum parts of the system at the quantum level (with the help of a Tight-Binding TB model) and embedding those parts in a semi-classical description
(valid on scales larger than the elastic mean free path).
This multi-scale approach is performed in three steps.  

First, we provide an extensive presentation of our semi-classical approach which we refer as Continuous Random Matrix Theory (CRMT). CRMT  was developed recently in Ref~\onlinecite{Rychkov:2009}
to which we refer  for a compact account of the theory. We discuss in details the links with other approaches 
(in particular the explicit mappings to circuit theory\cite{Brataas:2006} and Valet-Fert theory~\cite{Valet:1993}) and explain how classical concepts like spin accumulation naturally appear in the formalism, even in the absence of local equilibrium.

Second, we study a toy model for a tunneling spin valve and show explicitly how to combine CRMT with the TB approach. In particular, 
we discuss in details the conditions for the CRMT approach to be valid by comparing the multi-scale approach with full quantum calculations of the TB model.  Third and last  we construct an effective TB model capable of describing diffusive magnetic multilayers. This TB model is solved at a purely quantum level and we find a close agreement between TB and CRMT, so that the same parametrization can be used for both.  The appendix describes a direct integration of CRMT equations for a magnetic layer with no spin texture, together with a discussion of the boundary conditions and role of Sharvin resistances.

\section{Continuous Random Matrix Theory (CRMT)}

This section is devoted to the CRMT semi-classical theory that describes the conducting electrons of a magnetic multilayer.
This section can be considered as a long version of the letter published in Ref~\onlinecite{Rychkov:2009} to which we add an intuitive derivation of the relation between Scattering matrices and spin accumulation, making use of the voltage probe concept~\cite{Buttiker:1986, Brouwer:1995, Brouwer:1997}. CRMT is a continuous version of a Random Matrix Theory (RMT~\cite{Beenakker:1997}) for non-collinear magnetic multilayers that was developed in Ref~\onlinecite{Waintal:2000}, which is itself based on the fully quantum Landauer-Buttiker scattering approach~\cite{Buttiker:1986}. For the sake of completeness, both are briefly reviewed below.

\subsection{Landauer Buttiker approach to magnetic multilayers}

\begin{figure}
\includegraphics[keepaspectratio,width=0.7\linewidth]{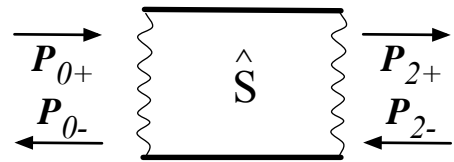}
\caption{\label{fig:S1} Cartoon of the S matrix approach. We define the region 0 on the left and 2 on the right (region 1 in between will appear later in the text). The + (-) respectively stands for right (left) going modes. The mode amplitudes are denoted ${\bm\psi}_{i\pm\sigma}$ at the quantum level and become probabilities ${\bm P}_{i\pm\sigma}$ in RMT. }
\end{figure}

The Landauer-Buttiker or scattering matrix formalism\cite{Landauer:1957,Landauer:1970,Buttiker:1985} is a standard approach to quantum transport. A sample is defined by the scattering matrix $S$ which expresses the outgoing propagating modes in term of the incoming ones.
The incoming states are filled according to the Fermi-Dirac distribution of the electrodes to which they are connected which enables to relate various physical quantities, such as conductance $G$ or spin current $\vec J$, to the matrix elements of $S$.
The system contains $N_{\rm ch}\gg 1$ propagating modes per spin where $N_{\rm ch}\approx A/\lambda_F^2$  ($A$ is the transverse area of the 
electrode and $\lambda_F$ is the Fermi wavelength). The amplitude of the wave-function on the different modes is given by a
vector ${\bm \psi}_{i\pm}$,
\be
{\bm \psi}_{i\pm}=\left(\begin{array}{c}{\bm\psi}_{i\pm\u} \\{\bm\psi}_{i\pm\d}\end{array}\right)
\ee
where ${\bm\psi}_{i\pm\sigma}$ is  a $N_{\rm ch}$ vector that contains the amplitudes for right (left) moving electron direction with spin $\sigma=\u,\d$ along the $z$-axis in region  $i=0,2$, see Fig.~\ref{fig:S1} for a cartoon.
With these definitions, the $S$ matrix is defined as,
\be
\label{eq:scattering-formalism}
\left(\begin{array}{c}{\bm\psi}_{0-} \\{\bm\psi}_{2+}\end{array}\right)=S\left(\begin{array}{c}{\bm\psi}_{0+}\\{\bm\psi}_{2-}\end{array}\right), \mbox{with} 
\ee
The $S$  matrix is a $4N_{\rm ch}\times4N_{\rm ch}$ unitary matrix and consists of $2N_{\rm ch}\times2N_{\rm ch}$ transmission $t,t'$ and reflection $r,r'$ sub-blocks,
\be
\label{eq:s}
S=\left(\begin{array}{cc}  r' & t \\ t' & r \end{array}\right),
\ee	
while the transmission and reflection matrices have an internal spin structure:
\be
t=\left(\begin{array}{cc}t_{\u\u} & t_{\u\d} \\ t_{\d\u} & t_{\d\d}\end{array}\right)
\ee
where $t_{\sigma\sigma'}$ are $N_{\rm ch}\times N_{\rm ch}$ matrices containing amplitudes for transmission between $\sigma'$ and $\sigma$ spin states.

The conductance of the system is given by the standard  Landauer formula $G=\frac{e^2}{h}\Tr \left[t^\dag t\right]$ while
the  spin current in region $0$ is given by,
\be
\label{eq:spin_current_0}
\frac{\partial\vec J_0}{\partial\mu}=\frac{1}{4\pi}\Tr\left[t\vec\sigma t^\dag\right]
\ee
where $\vec\sigma$ is the vector of Pauli matrices $\vec\sigma=(\vec\sigma_x,\vec\sigma_y,\vec\sigma_z)$ and $\mu$ the difference of chemical potential between the two electrodes (with similar formulas for the spin current inside the system~\cite{Waintal:2000}). Note that in addition to the non-equilibrium spin current, there can exist an equilibrium one which in the context of magnetic multilayers is often referred as the interlayer exchange (RKKY type) coupling\cite{Bruno:1992, Bruno:1995, Waintal:2002, Heinrich:2003}. 

\subsection{Random Matrix Theory (RMT) for magnetic multilayers}

So far, the theory is fully quantum and contains in particular many interference effects such as weak localization~\cite{Anderson:1979,Gorkov:1979} or universal conductance fluctuations~\cite{Altshuler:1985,Lee:1985}. We now proceed with the description of the semi-classical RMT\cite{Waintal:2000} theory. We give a presentation slightly more general and compact than in Ref.~\onlinecite{Waintal:2000} to which we refer for the proofs.
The systems in which we are interested in this paper, typically pillars of a few tens nanometers in diameter connected to top and bottom electrodes, contain many channels, typically $N_{\rm ch}\approx 10^4-10^5$. When the scattering is not perfectly ballistic (Fermi momentum mismatch at the interfaces, surface roughness or impurity scattering) those channels get mixed up. RMT assumes that this mixing is ergodic~\cite{Beenakker:1997}, (i.e an electron entering the system in a given mode will leave it in an arbitrary mode and pickup a random phase in the process). The theory can be derived systematically when the system size is larger than the mean free path, but we shall see that in practice, its domain of applicability is even wider.

RMT is expressed in term of $4\times4$ "hat" matrices that are obtained by tracing out the transverse degrees of freedom of the original reflection and transmission matrices. For instance, $\hat t$ is defined as,
\be
\label{eq:hat}
\hat t_{\sigma\eta,\sigma'\eta'}=\frac{1}{N_{\rm ch}}\Tr_{N_{\rm ch}}[t_{\sigma\sigma'} t^\dag_{\eta\eta'}],
\ee
($\Tr_{N_{\rm ch}}$ trace is taken on the modes only) or equivalently, giving the $4\times 4$ structure explicitly,
\be
\label{eq:t-hat} 
\hat{t}=\frac{1}{N_{\rm ch}} \Tr_{N_{\rm ch}}
\left(
\begin{array}{cccc} t_{\uparrow \uparrow} t_{\uparrow \uparrow}^{\dagger} &
t_{\uparrow \uparrow}t_{\uparrow \downarrow}^{\dagger} &
 		    t_{\uparrow \downarrow}t_{\uparrow \uparrow}^{\dagger} &
t_{\uparrow \downarrow}t_{\uparrow \downarrow}^{\dagger} \\
 		    t_{\uparrow \uparrow}t_{\downarrow \uparrow}^{\dagger} &
t_{\uparrow \uparrow}t_{\downarrow \downarrow}^{\dagger} &
 		    t_{\uparrow \downarrow}t_{\downarrow \uparrow}^{\dagger} &
t_{\uparrow \downarrow}t_{\downarrow \downarrow}^{\dagger} \\
                    t_{\downarrow \uparrow}t_{\uparrow \uparrow}^{\dagger}
&  t_{\downarrow \uparrow}t_{\uparrow \downarrow}^{\dagger} &
 		    t_{\downarrow \downarrow}t_{\uparrow \uparrow}^{\dagger} &
t_{\downarrow \downarrow}t_{\uparrow \downarrow}^{\dagger} \\
                    t_{\downarrow \uparrow}t_{\downarrow
\uparrow}^{\dagger} &  t_{\downarrow \uparrow}t_{\downarrow
\downarrow}^{\dagger} &
 		    t_{\downarrow \downarrow}t_{\downarrow \uparrow}^{\dagger} &
t_{\downarrow \downarrow}t_{\downarrow \downarrow}^{\dagger}
\end{array} \right).
\ee
Similarly,   \h-matrix $\hat S$ has a form similar to Eq.~(\ref {eq:s}),
\be
\label{eq:hat_s}
\hat S=
\left(\begin{array}{cc}  \hat r' & \hat t \\ \hat t' & \hat r \end{array}\right),
\ee	
In analogy with "hat" matrices, we introduce $4$-vectors  ${\bm P}_{i\pm}$which corresponds to the modes amplitudes ${\bm\psi}_{i\pm\sigma}$: 
\be 
\label{eq:P_vector}
{\bm P}_{i\pm}=\left(\begin{array}{l}{P}_{i\pm,\u}\\{P}_{i\pm, mx}\\{P}^{\ast}_{i\pm, mx}\\{P}_{i\pm,\d}\\\end{array}\right)
\ee
The components of the $4$-vector ${\bm P}_{i\pm}$  have interpretation in term of probabilities. For instance, 
${\bm P}_{0+\u}$ (${\bm P}_{2-\d}$) accounts for the probability to find a right (left) moving electron in region 0 (2) with spin $\u$ ($\d$).  The mixing components, ${P}_{\rm mx}$ are complex numbers which correspond to probability to find the electron along the $x$ (real part) or $y$ (imaginary part) axis. Inside magnetic layers where the $z$ axis will correspond to the direction of the magnetization, they will correspond to the (small) probability for the spin to have a part transverse to the magnetization.
Again, in analogy with Eq.~(\ref{eq:s}) which expresses the amplitudes of the outgoing modes in term of the incoming ones, we have,
\be
\label{eq:hat_scattering}
\left(\begin{array}{c}{\bm P}_{0-}\\{\bm P}_{2+}\end{array}\right)=\hat S\left(\begin{array}{c}{\bm P}_{0+}\\{\bm P}_{2-}\end{array}\right),
\ee
When the "mixing" components of the $4$-vectors play no role, as in collinear systems, Eq.~\ref{eq:hat_scattering} has an obvious interpretation in term of a Master equation. For instance, its first row expresses that the probability to find a left going electron in region $0$ has two contributions coming from the probability to have a reflection and transmission event. In other word, instead of the original interference problem with amplitudes, one now deals with the classical equivalent with probabilities. For non-collinear system however, the presence of the quantum SU(2) structure of the spin introduces some complex numbers (the mixing coefficients) and the Markov process interpretation does not hold, strictly speaking.

{\it Physical observables.} The chief result of Ref.~\onlinecite{Waintal:2000} are the expression for the currents and spin current in term of the "hat" matrices, to leading order in $N_{\rm ch}$. Equivalently, we can write those expression with the  ${\bm P}$-vectors,
\be
\label{eq:spin_current_P}
\vec J_i=\frac{N_{\rm ch}}{4\pi}\left[\vec{\bm \sigma}\cdot {\bm P}_{i+}-\vec{\bm\sigma}\cdot{\bm P}_{i-}\right]
\ee
and 
\be
\label{eq:current_P}
I=\frac{1}{e{\cal R}_{sh}}[{P}_{+\u}+{P}_{+\d} -{P}_{-\u}-{P}_{-\d}]
\ee

where $I$ is the charge current and ${\cal R}_{sh}=h/(N_{\rm ch}e^2)$ is the Sharvin resistance. $\vec{\bm \sigma}=(\vec\sigma_{\u\u},\vec\sigma_{\d\u},\vec\sigma_{\u\d},\vec\sigma_{\d\d})$ is $4$-vector composed of components of Pauli matrices.
To complete the theory, we need the boundary conditions imposed on the incoming electrons on both sides of the system.
We focus here on normal electrodes, but these conditions can be easily generalized to the case where the electrodes are themselves magnetic. We have,
\begin{eqnarray}
\label{eq:boundary_conditions}
{\bm P}_{0+}=\left(\begin{array}{l}\mu_0\\0\\0\\ \mu_0\end{array}\right),\ 
{\bm P}_{2-}=\left(\begin{array}{l}\mu_2\\0\\0\\ \mu_2\end{array}\right)
\end{eqnarray}
where $\mu_0$ and $\mu_2$ are the respective chemical potentials of the two electrodes.
For a given \h-matrix $\hat S$, the combinations of Eqs.~(\ref{eq:hat_scattering},\ref{eq:spin_current_P},\ref{eq:current_P}) and Eq.~(\ref{eq:boundary_conditions}) form a complete set of equations to obtain the physical quantities.

\begin{figure}
\includegraphics[keepaspectratio,width=0.7\linewidth]{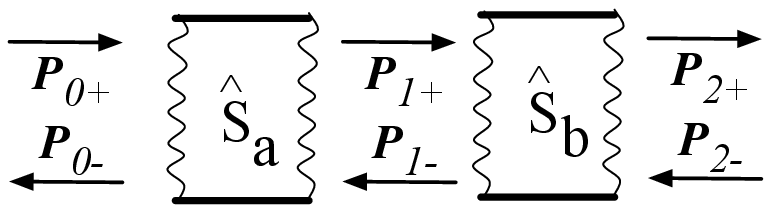}
\caption{\label{fig:S2}} Cartoon of a system made of two subsystems $a$ and $b$. 
\end{figure}

{\it Two systems in series.} Usually, the $\hat S$ matrix can be obtained for parts of the system (interfaces or the bulk part of one material) and it is important to be able to combine several parts to obtain the global $\hat S$ matrix. This is particularly important when one is interested in calculating the spin torque, as one also needs to calculate the spin current flowing {\it inside} the system and not only near the electrodes. Let us consider the transport through  a system which consists of two subparts, as pictured in Fig.~\ref{fig:S2}. The transport through each subpart is characterized by the corresponding \h-matrix $\hat S_a$ and $\hat S_b$. Eq.~(\ref{eq:hat_scattering}) reads for each of them,
\begin{eqnarray}
\label{eq:hat_scattering2}
\left(\begin{array}{c}{\bm P}_{0-}\\{\bm P}_{1+}\end{array}\right)=\hat S_a\left(\begin{array}{c}{\bm P}_{0+}\\{\bm P}_{1-}\end{array}\right),
\left(\begin{array}{c}{\bm P}_{1-}\\{\bm P}_{2+}\end{array}\right)=\hat S_b\left(\begin{array}{c}{\bm P}_{1+}\\{\bm P}_{2-}\end{array}\right)
\end{eqnarray}
Eliminating ${\bm P}_1$ we find addition law for \h-matrices:
\begin{eqnarray}
\label{eq:Sab}
\left(\begin{array}{c}{\bm P}_{0-}\\{\bm P}_{2+}\end{array}\right)&=&\hat S_{a+b}\left(\begin{array}{c}{\bm P}_{0+}\\{\bm P}_{2-}
\end{array}\right)
{\rm with, }\\
\label{eq:ht12}
\hat t_{a+b}&=&\hat t_a\frac{1}{\hat 1-\hat r'_b\hat r_a}\hat t_b\\
\label{eq:hr12}
\hat r_{a+b}&=&  \hat r_b+\hat t_b'\frac{1}{\hat 1-\hat r_a\hat r'_b}\hat r_a\hat t_b  
\end{eqnarray}
and similar expressions for $\hat r'_{a+b}$ and $\hat t'_{a+b}$.
We can also use Eq.~(\ref{eq:hat_scattering2}) to express ${\bm P}_{1\pm}$ in terms of the incoming fluxes: 
\be
\label{eq:hat_Pinside}
\left(\begin{array}{c}{\bm P}_{1+}\\{\bm P}_{1-}\end{array}\right)=
\left(\begin{array}{cc}
\frac{1}{\hat 1-\hat r_a\hat r'_b}\hat t'_a & \frac{1}{\hat 1-\hat r_a\hat r'_b}\hat r_a\hat t_b \\
\frac{1}{\hat 1-\hat r'_b\hat r_a}\hat r'_b\hat t'_a & \frac{1}{\hat1-\hat r'_b\hat r_a}\hat t_b
\end{array}\right)
\left(\begin{array}{c}{\bm P}_{0+}\\{\bm P}_{2-}\end{array}\right)
\ee
Using Eq.~(\ref{eq:spin_current_P}), Eq.~(\ref{eq:boundary_conditions}) and Eq.~(\ref{eq:hat_Pinside}), one can calculate the spin current in the region $1$ inside the system.

{\it Rotation matrices. } The scattering matrix of a piece of magnetic material is most easily found in the basis parallel to the local magnetization, i.e. the reflection and transmission matrices are given for the majority and minority electrons. For non-collinear multilayers where different magnetization directions come into play, one needs to rotate the original $S$ matrix onto its form $\tilde S= R_{\theta,\vec n}S R_{\theta,\vec n}^\dagger$ in the chosen working basis. Here $R_{\theta,\vec n}=\exp(-i\vec\sigma\cdot\vec n \ \theta/2)=\cos(\theta/2)-i\vec\sigma\cdot\vec n\ \sin(\theta/2)$ is the rotation matrix of angle $\theta$ around the unit vector $\vec n$ that brings the magnetization onto the z-axis of the working basis. In term of "hat" matrices, this translate directly into $\hat{\tilde S}= \hat R_{\theta,\vec n}\hat S \hat R_{\theta,\vec n}^\dagger$
with,
\be
\label{rotation}
\hat R_{\sigma\eta,\sigma'\eta'}=R_{\sigma\sigma'} R^{\ast}_{\eta\eta'},
\ee
a unitary matrix.

\subsection{From scattering degrees of freedom to spin accumulation and spin current}

The natural variables of RMT as it was introduced above are the $4$-vectors ${\bm P}_{x\pm}$ which characterize the "probability" to find a left or right moving electron in region $x$. Let us now introduce a new set of variables defined as,
\be
\label{eq:j}
{\mathbf j}(x) =  [{\mathbf P}_+(x) - {\mathbf P}_-(x)]/(e{\cal R}_{\rm sh}), 
\ee
\be
\label{eq:mu}
{\bm \mu}(x) = [{\mathbf P}_+(x) + {\mathbf P}_-(x)]/2
\ee
As we shall see, these two $4$-vectors correspond respectively to the (spin resolved) current and chemical potentials flowing in the system. There are two complementary ways to make this connection. The first one is to write our fundamental equations (essentially Eq.~(\ref{eq:hat_scattering})) in term of these new variables. This will be done toward the end of this section, and we will find that 
${\mathbf j}(x)$ and ${\bm \mu}(x)$ satisfies the well known Valet-Fert diffusive equations~\cite{Valet:1993} for continuous collinear systems and 
the equations of circuit theory for discrete non-collinear ones~\cite{Bauer:2003b}. This connection is very interesting from the theoretical point of view as very different routes have been taken to obtain these equations (scattering and Random matrix theory on one side and Keldysh Green function formalism and quasi-classical approximation on the other side). It has also a practical interest as one can use Eq.~(\ref{eq:j}) and Eq.~(\ref{eq:mu}) to go back and forth between a "scattering" approach and a "diffusive" approach and both have technical advantages for practical calculations.

There is a second, more direct, way to connect  respectively ${\mathbf j}(x)$ and ${\bm \mu}(x)$ to the concepts of spin currents and spin accumulation. For ${\mathbf j}(x)$, the connection is straightforward, as  Eq.~(\ref{eq:spin_current_P}) reads,
\be
\label{eq:spin_current_Pbis}
\vec J(x)=\frac{\hbar}{2e}\vec{\bm \sigma}\cdot {\mathbf j}(x)
\ee
while
\be
I={\mathbf j}_\u + {\mathbf j}_\d
\ee
so that ${\mathbf j}_\u$ and ${\mathbf j}_\d$ can be directly interpreted as spin currents for up and down electrons (while 
${\mathbf j}_{mx}$ accounts or spin current transverse to the z-axis).
The connection between ${\bm \mu}(x)$ and a hypothetical spin resolved chemical potential is more problematic as 
scattering theory does not have any notion of local chemical potential. In fact, the existence of a local chemical potential would imply some sort of local equilibrium inside the system. However, the theory that we have developed so far is purely elastic and such a local equilibrium is not present. We will find that everything happens {\it as if} there were some sort of local equilibrium, except for one important point: the presence or absence of the contact resistance. 

\begin{figure}
\includegraphics[keepaspectratio,width=0.8\linewidth]{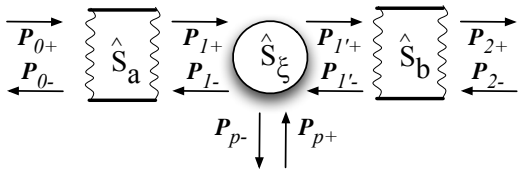}
\caption{\label{fig:voltage_probe} (a) Cartoon of the system in presence of the voltage probe.}
\end{figure}

 In order to provide a physical understanding of $\mu (x)$, we make use of a theoretical trick known as the voltage probe\cite{Buttiker:1986, Brouwer:1995, Brouwer:1997}: we connect a point inside the system (region $1$) to an external electrode and adjust the (spin resolved) chemical potential of this electrode so that no (spin) current flows from/to it. In the absence of magnetism, this could be achieved experimentally with the help of a STM tip. In mesoscopic physics, the voltage probe is used to induce some decoherence~\cite{Joos:book} in an otherwise perfectly coherent theory, hence introducing a finite phase coherence time which depends on the coupling of the system with the probe. Here, we will take this coupling to be extremely small, so that the probe will not affect the physics of the system. On the other hand, we are interested in the voltage that one must apply on the probe to stop any current from flowing from/to it: this will be our definition of an {\it effective} local chemical potential, and we shall find that this definition matches the one of $\mu (x)$. The setup is shown in Fig.~\ref{fig:voltage_probe}. In addition to $\hat S_a$ and $\hat S_b$ defined previously, we introduce the matrix $\hat S_\epsilon$ which describes the (small) coupling to the probe,
\be
\label{eq:beamspl}
\hat S_\epsilon=
\left(\begin{array}{ccc}
0          & (1-\epsilon) \, \hat 1& \epsilon\,\hat 1 \\
(1-\epsilon) \, \hat 1 & 0          & \epsilon \,  \hat 1\\ 
\epsilon \, \hat 1  & \epsilon \,  \hat 1  & (1-2\epsilon) \, \hat 1
\end{array}\right), 
\ee
with $\epsilon\ll 1$. When $\epsilon=0$ the probe is decoupled from the conductor and the electrons freely propagate between the regions $1$ and $1'$.%
The full system is entirely determined by the following set of equations,
\begin{eqnarray}
\label{eq:hat_probe_scattering}
&&\left(\begin{array}{c}{\bm P}_{0-}\\{\bm P}_{1+}\end{array}\right)=\hat S_a\left(\begin{array}{c}{\bm P}_{0+}\\{\bm P}_{1-}\end{array}\right),
\left(\begin{array}{c}{\bm P}_{1'-}\\{\bm P}_{2+}\end{array}\right)=\hat S_b\left(\begin{array}{c}{\bm P}_{1'+}\\{\bm P}_{2-}\end{array}\right)\nonumber\\
&&\left(\begin{array}{l}{\bm P}_{1-}\\{\bm P}_{1'+}\\{\bm P}_{p-}\end{array}\right)=\hat S_\epsilon \left(\begin{array}{l}{\bm P}_{1+}\\{\bm P}_{1'-}\\{\bm P}_{p+}\end{array}\right), 
\end{eqnarray}
to which one must add the incoming boundary conditions on the three electrodes ${\bm P}_{0+}$, ${\bm P}_{2-}$
[Eq.~(\ref{eq:boundary_conditions})] and ${\bm P}_{p+}$ (the probe).
Eliminating ${\bm P}_{1,1'}$ we obtain $\hat S$. In particular we get to leading order in $\epsilon$,
\begin{eqnarray}
\label{eq:vp1}
\hat S_{p,0}&=&\epsilon (\hat 1+\hat r'_b)\frac{1}{\hat 1-\hat r_a\hat r'_b}\hat t'_a\nonumber \\
\hat S_{p,2}&=&\epsilon (\hat 1+\hat r_a)\frac{1}{\hat 1-\hat r'_b\hat r_a}\hat t_b\nonumber \\
\hat S_{p,p}&=&1-2\epsilon\hat 1 + O(\epsilon^2)
\end{eqnarray}
We now impose that $\vec J_p=0$ [Eq.~(\ref{eq:spin_current_P})] which provides our boundary condition on the
probe:
\be
\label{eq:spin_acc}
{\bm P}_{p+}=\frac{1}{2\epsilon}\hat S_{p,0} {\bm P}_{0+}+\frac{1}{2\epsilon}\hat S_{p,2} {\bm P}_{2-}
\ee
Using Eq.~(\ref{eq:hat_Pinside}) and Eq.~(\ref{eq:mu}), we arrive at,
\be
\label{eq:chem_pot}
{\bm P}_{p+}=\frac{1}{2}\left( {\bm P}_{1+}+{\bm P}_{1-} \right)={\bm \mu}_1
\ee
In other words, the (spin resolved) voltages that one need to apply on the probe is equal (as announced above) to 
the formal chemical potential that we have defined in Eq.~(\ref{eq:mu}) so that it is legitimate to call $\mu (x)$ a chemical potential.

\subsection{Continuous Random Matrix Theory (CRMT)}

The basic ingredient of the RMT theory presented above are the "hat" scattering matrices of the various sub parts of the system. Once those are known, they can be concatenated to obtain various physical observables. There are two routes that one can take to obtain those "hat" matrix. The first one is to go back to their definition in term of the original quantum problem, and compute them from purely quantum calculations. This route has been taken with ab-initio calculation for the interfaces between several magnetic and non-magnetic layers\cite{Xia:2006}.
In the same spirit, in the next section, we will calculate the "hat" matrices from calculations of an effective quantum model for diffusive magnetic metals. In the second route, that we take here, we derive those "hat" matrices from phenomenological considerations. The $\hat S$ matrix for a piece of bulk material or for an interface between two different metals will be parametrized by a few parameters. These parameters will in turn be put in one to one correspondence with the parameters of the (well established) Valet-Fert theory\cite{Valet:1993}. This has a double advantage as it allows to make a direct connection with a widely spread theory, and it also allows to make direct use of the important experimental effort that has been done to parametrize Valet-Fert theory.

{\it General form of the "hat" matrices in a magnetic material.}  A typical magnetic multilayer consists of alternating layers of magnetic and non-magnetic materials of various widths. We decompose the corresponding $S$ matrices in the bulk and interface parts that will be parametrized independently.  The general form of a "hat" matrix ($\hat t$ in what follows but similar considerations apply to $t'$, $r$ and $r'$) is a full $4\times 4$ matrix given by  Eq.~(\ref{eq:t-hat}). 
In the absence of spin-orbit scattering (and/or magnetic impurities), spin is a good quantum number of the problem and the $t$ matrix is diagonal in spin space (in the basis parallel with the magnetization). Hence $\hat t$ is also diagonal. It consists on one hand of the probabilities $T_{\u\u}$  ($T_{\d\d}$) for an $\u$ ($\d$) spin to be transmitted and on the other hand of the so-called "mixing transmission", $T_{\rm mx}=(1/N_{\rm ch}) {\rm Tr_{N_{\rm ch}}}( t_{\uparrow \uparrow}t_{\downarrow \downarrow}^{\dagger})$.
The latter is a complex number. Its amplitude measures how much of a spin transverse to the magnetic layer can be transmitted through the
system while its phase amounts for the corresponding precession~\cite{Stiles:2002b}.  Due to large number of modes the average $T_{\rm mx}$ decays rapidly with the size of a ferromagnet.\cite{Stiles:2002b,Xia:2006} 
Thus the mixing terms are small in magnetic systems, but can (and do) play a role in non-collinear configurations nevertheless. 
Spin-orbit interaction however cannot be entirely neglected, as it leads to a finite spin diffusion length in the system. Hence, a finite probability $T_{\u\d}$ ($T_{\d\u}$) for a $\d$ ($\u$) spin to be transmitted as a $\u$ ($\d$) spin must be considered. 
In the following, we suppose that mixing elements (complex numbers) that also involve spin-flip scattering can be totally ignored, leading to the following form of the \h-matrix:
\be
\label{eq:CRMT-t-hat}
\hat{t}=\left(
\begin{array}{cccc} T_{\uparrow \uparrow} &0 & 0& T_{\uparrow \downarrow} \\
 		   0 & T_{\rm mx} & 0 & 0 \\
                   0 &  0 & T_{\rm mx}^{\ast} & 0 \\
                    T_{\downarrow\uparrow} &  0&0 & T_{\downarrow \downarrow}\end{array} \right).
\ee
with the probabilities $T_{\sigma\sigma'}=1/N_{\rm ch}\Tr_{N_{\rm ch}}(\hat t^\dag_{\sigma\sigma'}t_{\sigma\sigma'})$. Note that the unitarity of the 
$S$ matrix imposes the following constraint on the mixing coefficient,
\be
\label{eq:int-mx}
\vert T_{\rm mx}\vert \leq\sqrt{T_{\u\u}T_{\d\d}} 
\ee

{\it General form of the "hat" matrices in a non-magnetic material.} In a non-magnetic material there is no preferred direction for the spin thus a \h-matrix of the non-magnetic metal must be invariant with respect to rotations, as defined in Eq.~(\ref{rotation}). One finds by inspection that there are two $4\times4$ matrices invariant with respect to rotation around an arbitrary axis:
\begin{equation}
\label{eq:inv}
\hat I_1=\left(
\begin{array}{cccc}
  1 &0   &0  &0 \\
  0 &1   &0   &0\\
  0 &0   &1   &0\\
  0 &0   &0   &1
\end{array}
\right),\ 
\hat I_2=\left(
\begin{array}{cccc}
  1 &0   &0  &1 \\
  0 &0   &0   &0\\
  0 &0   &0   &0\\
  1 &0   &0   &1
\end{array}
\right)
\end{equation}
Hence, we write the "hat" matrix of a normal layer  as a combination of these two invariants:
\be
\label{eq:hat_normal}
\hat t= [T-2T_{\rm sf}] \hat I_1+ T_{\rm sf} \hat I_2.
\ee
where $T$ is the total transmission probability while $T_{\rm sf}$ is the probability of transmission with spin-flip.

{\it Obtaining $\hat S$ for bulk materials.} So far we have discussed the form of the "hat" matrices for an arbitrary subpart of the system. This general form is directly useful for the parametrization of the "hat" matrices of interfaces. For bulk material, we introduce the $\hat S$ matrix of a very thin slice of material of width  $\delta L$. $\hat S(\delta L)$ (and therefore the bulk properties of the material) is entirely characterized by
two matrices $\Lambda^t$ and $\Lambda^r$ defined as,
\be
\label{eq:thin}
\hat t(\delta L) = 1 - \Lambda^t \delta L  \ \ , \ \ \hat r(\delta L) = \Lambda^r \delta L
\ee
Once $\hat S(\delta L)$ is known, one can make use of the addition law derived in Eq.~(\ref{eq:ht12}) and Eq.~(\ref{eq:hr12})
to obtain a differential equation that allows the computation of $\hat S(L)$: combining $\hat S(L)$ and $\hat S(\delta L)$ one obtains
$\hat S(L+\delta L)$ which by taking the limit $\delta L\rightarrow 0$ provides,   
\begin{eqnarray}
\label{drdl}
\frac{\partial\hat r}{\partial L} &=& \Lambda^r -\Lambda^t\hat r -\hat r\Lambda^t 
+\hat r\Lambda^r\hat r \\
\label{dtdl}
\frac{\partial\hat t}{\partial L} &=& -\Lambda^t\hat t + \hat r\Lambda^r t
\end{eqnarray}
Eq.~(\ref{drdl}) and Eq.~(\ref{dtdl})
can be integrated to obtain   $\hat S(L)$ of an arbitrary  bulk part (see the Appendix). However, for numerical purposes, it is more efficient to use directly Eq.~(\ref{eq:ht12}) and Eq.~(\ref{eq:hr12}) which leads to an extremely fast integration time $\propto \log L$ (In practice, one starts with an extremely small piece of material described by Eq.(\ref{eq:thin})
and recursively doubles it size using Eqs.~(\ref{eq:ht12},\ref{eq:hr12}) until obtaining the full width $L$ of the layer).

The matrices $\Lambda^t$ and $\Lambda^r$ are parametrized by the general form given in Eq.~(\ref{eq:CRMT-t-hat}) an Eq.~(\ref{eq:hat_normal}). The parametrization is further constrained by the unitarity of $S$ (conservation of current). Due to the origin essentially ballistic of the finite mixing coefficient, we neglect the corresponding contribution to $\Lambda^r$.
Eventually, a bulk magnetic material is characterized by four independent parameters $\Gamma_\uparrow$, $\Gamma_\downarrow$ $\Gamma_{\rm sf}$ and $\Gamma_{\rm mx}$:
\begin{eqnarray}
\label{eq:lambda}
\Lambda^t &=&
\left(\begin{array}{cccc}  
\Gamma_\uparrow +\Gamma_{\rm sf} & 0              &  0             & -\Gamma_{\rm sf} \\ 
           0                     & \Gamma_{\rm mx}&  0             & 0 \\
           0                     & 0              &\Gamma^{\ast}_{\rm mx} & 0 \\
        -\Gamma_{\rm sf}         & 0              & 0              & \Gamma_\downarrow+\Gamma_{\rm sf} 
\end{array}\right),
\\
\label{eq:lambda_r}
\Lambda^r &=&
\left(\begin{array}{cccc}  
\Gamma_\uparrow -\Gamma_{\rm sf} & 0              &   0             & \Gamma_{\rm sf} \\ 
           0                                               & 0              &   0             & 0 \\
           0                                               & 0              &   0             & 0 \\
        \Gamma_{\rm sf}                        & 0               &   0             & \Gamma_\downarrow-\Gamma_{\rm sf} 
\end{array}\right),
\end{eqnarray}

These four parameters correspond in turn to 5 different lengths. 
The two most important one are the mean free paths for majority ($l_\u$) and minority ($l_\d$) electrons defined as
$l_\sigma = 1/\Gamma_\sigma$. Next comes the spin diffusion length $l_{\rm sf}=[4\Gamma_{\rm sf} (\Gamma_\u +\Gamma_\d)]^{-1/2}$ 
 (see below Eq.~(\ref{eq:C-RMT-VF}) and the Appendix).
Last the complex number $\Gamma_{\rm mx}= 1/l_\perp+i/l_{\rm L}$ where $l_\perp$ is the penetration length of transverse spin current inside the magnet while $l_{\rm L}$ is the Larmor precession length. These definitions are supported by the form of $T_{\rm mx}(L)$ which is readily obtained by integrating Eq.~(\ref{dtdl}),
\be
T_{\rm mx}(L)=e^{-L/l_\perp -iL/l_{\rm L}}
\ee
$l_\perp$ and $l_{\rm L}$,  are believed to be roughly equal, and the smallest characteristic lengths with typical values in the $nm$ range.
The explicit form of $T_{\sigma\sigma'}(L)$ will be discussed in the appendix.

In a normal metal, the parametrization obeys Eq.~(\ref{eq:hat_normal}) and we have
 \be
\Lambda^t=[\Gamma+2\Gamma_{\rm sf}] \hat I_1- \Gamma_{\rm sf} \hat I_2.
\ee
\be
\Lambda^r=[\Gamma-2\Gamma_{\rm sf}] \hat I_1+ \Gamma_{\rm sf} \hat I_2.
\ee
The theory is now formally complete. A given multilayer is then constructed 
by using the addition law Eqs. (\ref{eq:ht12},\ref{eq:hr12}) for the various bulk layers and 
the corresponding interfaces. 

\subsection{Correspondence between CRMT and Valet Fert theory}\label{SS:link_to_VF}

We are now ready to connect CRMT with other approaches taken in the literature. An important connection that we make here is to show that for collinear systems, CRMT equations are equivalent to the celebrated Valet-Fert (VF) equations.\cite{Valet:1993}
As these equations have been extensively parametrized with the huge corpus of experimental data available in CPP GMR, this allows us to transfer directly this parametrization to CRMT.

{\it Bulk magnetic materials.} Let us start with $\hat S(\delta x)$ which relates ${\mathbf  P}_{\pm}(x)$ and ${\mathbf  P}_{\pm}(x+\delta x)$
on the two sides of a thin slice of material according to Eq.~(\ref{eq:hat_scattering}). The explicit form of  $\hat S(\delta x)$ given by Eq.~(\ref{eq:thin}) provides  differential equation for ${\mathbf  P}_{\pm}(x)$ which are the counterparts of Eq.~(\ref{drdl}) and Eq.~(\ref{dtdl}):
\begin{eqnarray}
\label{eq:diff}
\frac{\partial{\bm P}_+(x)}{\partial x}&=& -\Lambda^t {\bm P}_+(x)+\Lambda^r {\bm P}_-(x)\nonumber\\
\frac{\partial{\bm P}_-(x)}{\partial x}&=& -\Lambda^r {\bm P}_+(x)+\Lambda^t {\bm P}_-(x).
\end{eqnarray}
In the absence of magnetic texture, the equations for $P_{\pm,\uparrow}$ and $P_{\pm,\downarrow}$ are decoupled from the equations for  $P_{\pm,mx}$ and $P^{\ast}_{\pm,mx}$.  Focussing on the  non-mixing contributions (first and last row of the 4-vectors) of   Eq.~(\ref{eq:diff}) we get explicitly,
\begin{eqnarray}
\label{eq:CRMT2VF}
\frac{\partial}{\partial x}\left(\begin{array}{c}{P}_{\u\pm}\\{P}_{\d\pm}\end{array}\right)&=& -\tilde\Lambda^t \left(\begin{array}{c}{P}_{\u\pm}\\{P}_{\d\pm}\end{array}\right)+\tilde\Lambda^r \left(\begin{array}{c}{P}_{\u\mp}\\{P}_{\d\mp}\end{array}\right),
\end{eqnarray}
where $\tilde\Lambda^{t}$ and $\tilde\Lambda^{r}$ are $2\times 2$ matrices,
\be
\label{eq:lambda_tilde}
\tilde\Lambda^{t/r}=\left(\begin{array}{cc}  
\Gamma_\uparrow \pm\Gamma_{\rm sf}              & \mp\Gamma_{\rm sf} \\ 
        \mp\Gamma_{\rm sf}                 & \Gamma_\downarrow\pm\Gamma_{\rm sf} 
\end{array}\right).
\ee
Eq.~(\ref{eq:CRMT2VF}) accounts for the conservation of probability in the scattering events, i.e. it is the Master equation of the underlying Brownian motion undertaken by the electrons upon the various reflection and transmission events.

Using Eq.~(\ref{eq:j}) and Eq.~(\ref{eq:mu}) we can now write Eq.~(\ref{eq:CRMT2VF})  in terms of ${\mathbf  j}(x)$ and ${\bm \mu}(x)$, 
and arrive at,
\begin{eqnarray}
\label{eq:vf}
j_{\u/\d}(x) &=& -1/(e\Gamma_\sigma{\cal R}_{\rm sh})\  \partial_x \mu_{\u/\d}(x) \\
\partial_x j_{\u/\d}(x) &=& 4 \Gamma_{\rm sf} / (e{\cal R}_{\rm sh})\  [\mu_{\d/\u}(x) - \mu_{\u/\d}(x)]
\end{eqnarray}
which are precisely the Valet-Fert equations~\cite{Valet:1993}. Hence, for a collinear system, CRMT simply reduces to
VF theory. This allows us to build a one to one correspondence between the CRMT parameters ($\Gamma_\u$, $\Gamma_\d$
and $\Gamma_{\rm sf}$) and the VF parameters [$\rho_\u$, $\rho_\d$ (resistivities for majority and minority electrons) and $l_{\rm sf}$
(spin-flip diffusion length)]. Using the standard notations for the average resistivity $\rho^{\ast}$ and polarization $\beta$ [$\rho_{\uparrow(\downarrow)}= 2\rho^{\ast} (1\mp \beta)]$ we find
\begin{eqnarray}
\label{eq:C-RMT-VF}
\frac{1}{l_{\rm sf}} &=& 2\sqrt{\Gamma_{\rm sf}}\sqrt{\Gamma_\uparrow + \Gamma_\downarrow} \label{eq:C-RMT-VF-lsf} \\
\beta&=&\frac{\Gamma_\downarrow - \Gamma_\uparrow}{\Gamma_\uparrow + \Gamma_\downarrow}\label{eq:C-RMT-VF-beta}\\
\frac{\rho^{\ast}}{{\cal R}_{\rm sh}}&=&(\Gamma_\uparrow+\Gamma_\downarrow)/4\label{eq:C-RMT-VF-rho*},
\end{eqnarray}
Note that the mixing coefficient $\Gamma_{\rm mx}$ is not fixed by this parametrization as they only play a role in non collinear configurations. So far there are very few experimental data\cite{Taniguchi:2008a, Taniguchi:2008b, Taniguchi:2008c} allowing to extract $\Gamma_{\rm mx}$ so that one often relies on model or
{\it ab-initio} calculations to estimate it.

{\it Interfaces.} In VF theory, the interfaces are characterized by effective interface
resistances $r_{\u}$ ($r_{\d}$) for majority (minority) electrons as well as the spin-flip probability $\delta$.
Alternatively one can introduce the average resistance $r^{b\ast}$ and polarization $\gamma$ [$r_{\u,\d}=2r^{b\ast} (1\mp \gamma)$]. Much is known experimentally\cite{Yang:1995,Xia:2001} about  $r^b_{\u,\d}$, but there are much less experimental data\cite{Park:2000} for $\delta$ (and those are mainly for normal-normal interfaces).
In VF theory, the interface boundary conditions are obtained by introducing an virtual material of width $d$, resistivity $\rho^{eff}_{\sigma}$ and spin diffusion length $l^{eff}_{\rm sf}$. This virtual material is then taken to be infinitely thin $d\rightarrow 0$ while keeping the interface parameters finite: $\delta=d/l^{eff}_{\rm sf}$ and $r^{b}_{\sigma}=\rho^{eff}_{\sigma} d$.
Repeating the same procedure for CRMT allows us to map the VF parameters to CRMT interfaces,
\begin{eqnarray}
\label{eq:interface}
&&T_{\u\u}=\frac{(1+e^{-\delta})/2}{1+2(r^{b\ast}/{\cal R}_{sh})(1-\gamma)}\\
&&T_{\d\u}=\frac{(1-e^{-\delta})/2}{1+2(r^{b\ast}/{\cal R}_{sh})(1-\gamma)}\\
&&T_{\u\d}=\frac{(1-e^{-\delta})/2}{1+2(r^{b\ast}/{\cal R}_{sh})(1+\gamma)}\\
&&T_{\d\d}=\frac{(1+e^{-\delta})/2}{1+2(r^{b\ast}/{\cal R}_{sh})(1+\gamma)}\\
&&R_{\u\u}=1-\frac{1}{1+2(r^{b\ast}/{\cal R}_{sh})(1-\gamma)}, R_{\d\u}=0 \\
&&R_{\d\d}=1-\frac{1}{1+2(r^{b\ast}/{\cal R}_{sh})(1+\gamma)}, R_{\u\d}=0 \\
\end{eqnarray}
Alternatively, the interface $\hat S$ matrix can be obtained directly from {\it ab-initio} or model quantum calculations (see the next section). Once again, the mixing reflection and transmission coefficients are not fixed by CPP GMR experiments. Those numbers can be extracted from angular resolved magneto-resistance or spin pumping experiments but limited data are available so far\cite{Tserkovnyak:2005}.

{\it Valet-Fert theory and Sharvin resistances.} We have just proved that for a collinear system, VF theory and CRMT are simply equivalent. There is however a small difference  that appears in the boundary conditions at the electrodes. In CRMT those boundary conditions come from  the (quantum) Landauer formula and the presence of different voltages between the reservoirs located at, say, $x=0$
and $x=L$ imposes $P_{+\sigma}(0)=eV(0)$ and $P_{-\sigma}(L)=eV(L)$ [see Eq.~(\ref{eq:boundary_conditions})].
A direct consequence of these boundary conditions is the existence of a finite resistance, even for perfectly transparent interfaces and materials with negligible resistivity. In the context of mesoscopic physics, this leads to the quantization of conductance in unit of $2e^2/h$ which has been observed repeatedly.\cite{Wees:1988, Wharam:1988}.
These boundary conditions can be expressed in term of $\mu_\sigma (x)$ and $j_\sigma(x)$, and give,
\begin{eqnarray}
\label{eq:bc}
{\mathbf \mu}_\sigma(0) +(e{\cal R}_{\rm sh}/2)\ {j}_\sigma(0)&=&eV(0) \\
{\mathbf \mu}_\sigma(L) - (e{\cal R}_{\rm sh}/2)\ {j}_\sigma(L)&=&eV(L) 
\end{eqnarray} 
In other words, one needs to add ${\cal R}_{\rm sh}/2$ resistors on the two sides of the multilayer. For typical spin valve pillars, the intrinsic resistance of the pillar is only a few time ${\cal R}_{\rm sh}$
so that one really needs to take into account the presence of these Sharvin resistances in series to properly describe the pillar. We note that those proper boundary conditions can easily be included in a standard VF calculation. A detailed discussion of the effect of ${\cal R}_{\rm sh}$ on the polarization of spin current will be discussed in the Appendix.

\subsection{Correspondence between RMT and (generalized) Circuit Theory}

Let us now turn to non-collinear configurations. An alternative popular and powerful approach 
in this case is the so-called
circuit theory\cite{Bauer:2003b} where the system is discretized into various parts connected by "nodes" where one defines the spin resolved chemical potential. Circuit theory, initially derived for very resistive elements has been further extended into the "generalized" circuit theory to properly take into account the Sharvin resistance. The similitude between (generalized) circuit theory and RMT was recognized very early and it was shown in many cases that both theories gave the same result\cite{Brataas:2006}. 
Here we show that the analogy is in fact complete: RMT and generalized circuit theory are the same theory written in different variables: ${\mathbf  P}_{+}$ and ${\mathbf  P}_{-}$ for RMT and $\bm \mu$ and $\mathbf  j$ for circuit theory. One simply goes from one to the other using Eq.~(\ref{eq:j}) and Eq.~(\ref{eq:mu}).

The most general version of RMT between the two sides of a conductor is given by Eq.~(\ref{eq:P_vector}).
Turning now to $\bm \mu$ and $\mathbf  j$ variables on the two sides $L$ (left) and $R$ (right) of the conductor, we get,
\begin{eqnarray}
\label{eq:GenGenCT1}
(\hat 1+\hat r+\hat o' \hat t)\frac{{e\cal R}_{\rm sh}}{2}{\mathbf  j}_R = 2\hat o' {\bm \mu}_L - (\hat 1-\hat r+\hat o' \hat t){\bm \mu}_R \\
\label{eq:GenGenCT2}
(\hat 1+\hat r'-\hat o \hat t')\frac{{e\cal R}_{\rm sh}}{2}{\mathbf  j}_L = (\hat 1-\hat r'+\hat o \hat t'){\bm \mu}_L- 2\hat o {\bm \mu}_R
\end{eqnarray}
where $\hat o=t (1+\hat r)^{-1}$ and $\hat o'=t' (1+\hat r')^{-1}$. These two equations can be seen as the generalization of Ohm law and their linear combination provides the conservation equation for spin current.
Eq.~(\ref{eq:GenGenCT1}) and Eq.~(\ref{eq:GenGenCT2}) can be considered as an extension of (generalized) circuit theory, including in particular interface spin-flip scattering.

To recover generalized circuit theory, we need to make a few assumption. First, we neglect spin-flip scattering so that the "hat" matrices are purely diagonal (in the local basis of the magnetization). Second, we
suppose that $R_{\rm mx}$ might be non zero but set $T_{\rm mx}=0$. Eq.~(\ref{eq:GenGenCT1}) and Eq.~(\ref{eq:GenGenCT2}) simplify into
\begin{eqnarray}
\label{eq:circuit1}
j_{L\sigma}=j_{R\sigma}&=&\frac{1}{{e\cal R}_{\rm sh}}
\frac{T_{\sigma\sigma}}{1-T_{\sigma\sigma}} [\mu_{L\sigma}-\mu_{R\sigma}]\\
\label{eq:circuit2}
j_{L/R,\rm mx} &=& \pm \frac{2}{{e\cal R}_{\rm sh}}
\frac{1-R_{\rm mx}^{L/R}}{1+R_{\rm mx}^{L/R}} \mu_{L/R,\rm mx}
\end{eqnarray}
where $R_{\rm mx}^{L/R}$ are the mixing reflections from left to left ($R_{\rm mx}^{L}$) and right to right ($R_{\rm mx}^{R}$). Eq.~(\ref{eq:circuit1}) and Eq.~(\ref{eq:circuit2}) are precisely the equations that define generalized circuit theory~\cite{Bauer:2003b} which proves the equivalence with (C)RMT in this limit. In fact, the renormalization coefficients of generalized circuit theory~\cite{Bauer:2003b} were chosen such that the calculation of the conductance with RMT and generalized circuit theory fully agree with each other.

\section{A multiscale approach to spin transport.}

CRMT theory is essentially a spin extension of Ohm law  and can be considered as an non collinear extension of the diffusion equation. Hence, the resistance of a system of length $L$ typically decreases as $1/L$ and the theory cannot describe  the exponential suppression of the resistance found in purely quantum effects such as Anderson localization or quantum tunneling.  On the other hand, these effects (especially the latter) can play an important role in actual devices,  such as  tunneling magneto-resistance (TMR~\cite{Moodera:1995,Miyazaki:1995}) based spin valves.  For those systems  we need to take a completely different approach from the CRMT theory described above and derive a full quantum mechanical approach.   Such an approach is {\it a priori} very tempting as it  allows to deal with many different systems (metals, tunnel junctions, magnetic semi-conductors...) within one unified framework and using only few approximations. However, for practical calculations, such a path has a prohibitive numerical cost so that only very small systems can be studied (even for a minimum model such as the one we will introduce in  Section~\ref{S:TB}). 

Hence, instead of a full quantum general treatment, we have a slightly less ambitious double goal: 

(i) we want to study the regime of validity of CRMT  as well as deviations coming from quantum effects.

(ii) we want a multiscale description of the system: we will provide a purely quantum description of the pieces that require it (for instance the insulating oxide layers) while the rest of the system is described by CRMT. The purely quantum calculation will provide
effective boundary conditions for CRMT that account for whatever quantum effect is going on there.

A  multiscale approach is rather natural to implement in the CRMT framework. Indeed, as the basic ingredient of the theory are the "hat" matrix, we only need   to   provide the "hat" matrix   corresponding to the purely quantum parts of the system.  
Those can  in principle be calculated through their microscopic definitions, Eq.~(\ref{eq:hat}).    This approach has been applied with good success with {\it ab-initio} calculations of {\bf FN}     interfaces~\cite{Sanvito:1999, Haney:2007a, Haney:2007b,Waldron:2007}.

In this section, we first show how  the "hat" matrices  can be  expressed through the intrinsic microscopic properties of the system encoded  in the  retarded Green function.  Second, we introduce a toy model for a ferromagnetic-insulating-ferromagnetic spin valve to provide a practical example of the multiscale approach and study its regime of applicability. Last, we discuss an effective TB model which captures the main features of diffusive ferromagnetic layers and compare it with pure CRMT calculations. Technically, all the quantum numerical calculations presented below were done using the KNIT package\cite{Kazymyrenko:2008, Rychkova:2010} within the non equilibrium Green function (NEGF) formalism~\cite{Caroli:1971,Meir:1992}.  We refer to Ref.~\onlinecite{Kazymyrenko:2008, Rychkova:2010} for technical details on the numerical algorithm.

\subsection{ Connection between Semiclassical RMT and  Quantum NEGF  } %

Within the   NEGF formalism,  the transport properties of a system are related to the Green function $G$ of the system which can be calculated numerically recursively.
On the other side the  "hat" matrices are obtained from  the microscopic quantum scattering matrix $S$, as given in Eq.~(\ref{eq:hat}).
In order to connect this two approaches we make use of the Fisher-Lee formula\cite{Fisher:1981} which  relates the scattering 
matrix $S$ to the retarded Green function $G$ and obtain an explicit formula for $\hat S$ in term of the Green functions of the system.

According to Fischer-Lee formula\cite{Fisher:1981},
 the transmission (reflection)  amplitude from a mode $m$ in the the spin state $\sigma '$ on the right lead (region $2$ in Fig.~\ref{fig:S1}) to  a mode $n$ in the spin state $\sigma $ on the left  (right) lead ( respectively region $0$ or $2$ in Fig.~\ref{fig:S1}) reads, 
\begin{eqnarray}
\label{eq:fischer_lee-t}
t_{\sigma\sigma'}^{nm} & = & - {\it i}\hbar\, \sqrt{v_{n}v_{m}}\sum_{ij}{\bar\chi_{\sigma i}^{n*}}G_{\sigma i,\sigma' j}\chi_{\sigma' j}^{m},\\ 
\label{eq:fischer_lee-r}
r_{\sigma\sigma'}^{nm} & = &\delta_{nm} \, \delta_{\sigma\sigma'} -{\it i}\hbar\,  \sqrt{ v_{n}v_{m}}\sum_{ij}{\chi_{\sigma i}^{n*}}G_{\sigma i,\sigma' j}\chi_{\sigma' j}^{m}
\end{eqnarray}
where   $v_{n}$  is the velocity of an electron in channels $n$. The summation is taken over interface sites $ij$ which are on the border between the system and a given  lead,  $\chi_{\sigma' j}^{m}$ ($\bar\chi_{\sigma' j}^{m}$) is the transverse wave function of the $m$-th mode with spin $\sigma'$ evaluated at  site $j$ on the Right (Left) lead and  finally $G_{\sigma i,\sigma' j}$ is the retarded Green function between the spin state $\sigma '$ at site $j$ and spin  $\sigma$ at site $i$.
Note that the size of the transmission/reflection matrices $2N_{\rm ch}\times 2N_{\rm ch}$ is smaller than the total number of transverse sites, see Ref.~\onlinecite{Kazymyrenko:2008} for the  definition of the retarded Green functions $G$. 

With these notations,  The imaginary part of the reservoirs self energy $\Gamma^{L}$ for the Left lead reads,
\begin{eqnarray}\label{eq:gamma}
\Gamma^{L}_{\eta'\sigma,kl}=\sum_n \hbar v_{n} {\bar\chi}^{n}_{\eta' k} {\bar\chi}_{\sigma l }^{n\ast} .
\end{eqnarray}
with a similar definition for $\Gamma^{R}$.

The "hat" matrices $\hat t$  is  obtained by inserting Eqs.(\ref{eq:fischer_lee-t},\ref{eq:gamma}) into  Eq.~(\ref{eq:t-hat}), that yields,
\begin{eqnarray}
\label{eq:TB-t-hat}
\hat t_{\sigma\eta,\sigma'\eta'}
= \frac{1}{N_{\rm ch}}
\Tr_{N_{\rm ch}}
\left( \Gamma^L_{\eta'\sigma}G_{\sigma\sigma'}  \Gamma^R_{\sigma'\eta}G_{\eta\eta'}^\dagger \right).
\end{eqnarray}
where matrix notation has been used for the summation over the site indices.
Following the same path with  Eq.~(\ref{eq:fischer_lee-r}), the "hat" matrices $\hat r$  read,
\begin{eqnarray}
\label{eq:TB-r-hat}
\hat r_{\sigma\eta,\sigma'\eta'}
&=&\delta_{\sigma\sigma'}\delta_{\eta\eta'}+
\frac{1}{N_{\rm ch}}
\Tr_{N_{\rm ch}}
\left( \Gamma^R_{\eta'\sigma}G_{\sigma\sigma'}  \Gamma^R_{\sigma'\eta}G_{\eta\eta'}^\dagger \right)
 \nonumber
 \\&+&
\! \frac{{\it i} }{N_{\rm ch}}
\Tr_{N_{\rm ch}}
\left( \delta_{\sigma\sigma'} \Gamma^R_{\eta\eta'} G_{\eta\eta'}^\dagger - \delta_{\eta\eta'}  \Gamma^R_{\sigma\sigma'} G_{\sigma\sigma'}\right)\qquad
\end{eqnarray}

Using Eqs.(\ref{eq:TB-t-hat},\ref{eq:TB-r-hat})  we can now translate the results  from a quantum Tight-Binding calculation into our CRMT framework  and  embed quantum TB calculations inside CRMT. Alternatively, it will also allow us to compare pure CRMT
with TB calculations and map  the  experimental Valet-Fert parameters $\rho^{\ast},\beta,l_{\rm sf}$ and $l_{\perp},l_{\rm mx}$  into our minimum  Tight-Binding model for diffusive metallic layers (see Section~\ref{SS:min_TB}).

\subsection{Toy model of a TMR spin valve .}

This section is devoted to studying the accuracy of the semi-classical approximation done in RMT and hence of the multi-scale approach advocated in this paper. 

The cornerstone of RMT is the "hat"  matrix addition law for two systems put in serie, as defined by Eqs.~(\ref{eq:ht12},\ref{eq:hr12}). Its validity is based on an ergodic  assumption (an electron entering the system in one channel will leave it in arbitrary one) and secondly on the system
having many channels so that most quantum effects (such as  weak localization and universal conductance fluctuations) can be safely neglected.  In term of length scale,  these two restrictions are related to the  mean free path $\ell$ and the localization length 
$\xi$.  On one side the size of our  system  $L$  must be  long enough ($L \gg \ell$) to reach  the diffusive  regime, and  therefore  suppress  the  ballistic effects that reduce the mixing between channels.  On the other side the system's size should be small compared to the localization length  ($L \ll \xi$)   in  order to avoid strong localization,  that is not captured by our semiclassical theory.   Within the TB framework, the  inequality $ \ell \ll L \ll \xi$  translates into a  window range  for  the strength $W$ of the 
onsite disorder seen by the electrons.  Indeed for the quasi one dimensional geometry considered here, $1/\ell\propto W^2$
and $\xi/\ell\propto N_{\rm ch}$. By increasing sufficiently  the number of open  channels,   we can be sure to satisfy both requirements.
 
If we now add to the system a tunnel barrier, most of the open channel will  endure  a strong reduction of their  transmission's  probability.   This corresponds  in practice to an effective reduction of  number of channel really involved.  This render the system more sensible to the influence interference effects (universal conductance fluctuations) and we shall see we will need an even larger number of channels to ensure a good agreement between RMT and TB calculations. In realistic magnetic nanopillars, 
typically  $10^4-10^5$ channels are involved so that RMT is very well justified. However, some care is needed in the TB calculations as only a small number of channels (a few hundreds) can be considered there.
To summarize, the  three main limitation on the applicability of the addition law of "hat" matrices are :

(a) {\it Randomization/Mixing} of transverse motion, that require a strong enough disorder, ($L \ll \ell$ ).

(b)  {\it Strong Localization},  a purely quantum effect  that will avoid the use of too strong disorder, ($L \gg \xi$).

(c)  {\it Quantum Filtering},  that reduces the effective number of  channel and increases the  sensibility to the mesoscopic fluctuations.

\begin{figure}
\includegraphics[keepaspectratio,width=0.6\linewidth]{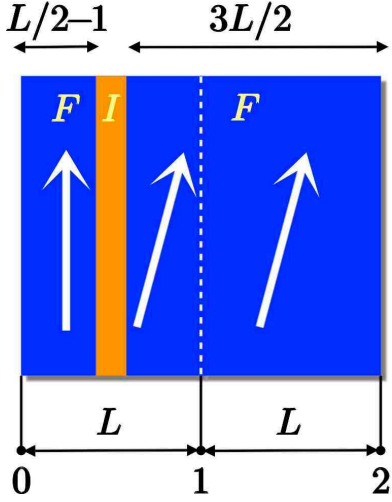}
\caption{\label{fig:FIF} Sketch of the Ferromagnetic-Insulator-Ferromagnetic ({\bf FIF}) spin valve used to study the validity of the
multi-scale approach.}
\end{figure}
We now consider a tunneling spin valve sketched Fig.~\ref{fig:FIF} and introduce a TB toy model for its description. 
The ferromagnetic ({\bf F}) and insulating layers  ({\bf I})  of the {\bf FIF}-system is  described with a minimalist TB-model with the  following  Hamiltonian:
\begin{equation}
\label{eq:mini-TB-H}
H=-{\sum_{<i,j>} {\bm c}_{i}^{\dagger} {\bm c}_j} +{\sum_i{\bm c}_{i}^{\dagger}{V}_i{{\bm  c}_i}} + {\sum_{i\in {\bm I}}{\bm c}_{i}^{\dagger}{U}_i{{\bm  c}_i}}
\end{equation}
Here the summation is  taken over nearest neighbor sites $<i,j>$ and $\bm I$ denote the set of sites included into the insulator. The electron destruction operator ${\bm c}_i$ is a spinor with components: ${\bm c}_i=(c_{i}^{\u},c_i^{\d})$ where $c_i^{\sigma}$ annihilates an electron with spin $\sigma$ on site $i$.  The  discretized  system is composed of $2L\times M\times M$ cells on a cubic lattice, where the $x$ axis is the direction of growth of the pillar.  The  onsite energy of TB-model is given by the spin dependent scattering potential.
\begin{equation}\label{eq:scat-pot}
V_i=v_i\left(
\begin{array}{cc} 
W^{\u} & 0\\
0 & W^{\d}
\end{array}\right),
\end{equation}
where the electrons with different spin $\sigma$ feel the same profile of the disorder defined by random numbers $v_i$ distributed according to a uniform
distribution with $v_i\in[-0.5,0.5]$, but with different strength $W^{\sigma}$.  In  this toy  model  the tunneling region contain an  insulator of  length of $1$ site  positioned after $L/2$ sites. The  insulating behavior   is described by the replacement of the onsite energy  by a spin dependent potential  $U_i$ 
 \begin{equation}\label{eq:ins-pot}
U_i= 
\left(
\begin{array}{cc} 
U^{\u}  & 0\\
0 & U^{\d}
\end{array}\right)
-V_i,
\end{equation}
In the following, we will compare RMT with pure TB calculations using two separate calculations: (i) a TB calculation of the full system of size $2L$ and (ii) We split the system into two parts of size $L$, calculate the hat matrices of the two sub part using
Eqs.(\ref{eq:TB-t-hat},\ref{eq:TB-r-hat}) and recombine them using the RMT "hat" addition law.

Before dealing with the full spin valve system, we need to establish  the parameters for which  the limitations (a,b)  are fulfilled. To do so, we  perform a spin independent survey ($W^{\sigma} = W$) in absence of the insulator ($U_i=0$ for all sites),  and  determine the  disorder's strength  range  for which  RMT agrees with the TB model . 
  In this particular case  the  case the "hat" matrix  reduce to scalar and addition  law  from transmission  simplifies  to, 
\begin{equation}
\label{eq:hat-add-nospin}
T_{\rm RMT} = \frac{ T_{\rm left} T_{\rm right }}{1-  R_{\rm left} R_{\rm right }}, 
\end{equation}
where $T_{\rm RMT} $ is the transmission probability  of the full system and $  T_{\rm left} $ ( $  T_{\rm right} $)  is  the transmission probability  of  the left (right) half of the system. 
 We  now compare the RMT result $ T_{\rm RMT} $ with  the one directly obtained by  $T_{\rm TB}$ calculation obtain for the whole system and compute the relative error  in percent between RMT and TB  predictions:  
 \be\label{eq:error}
 E\% =\frac{ 100\,\vert T_{\rm RMT} -T_{\rm TB}\vert}{ {1 \over 2}( T_{\rm RMT} +T_{\rm TB}) } 
 \ee
 
 \begin{figure}
\includegraphics[keepaspectratio,width=0.9\linewidth]{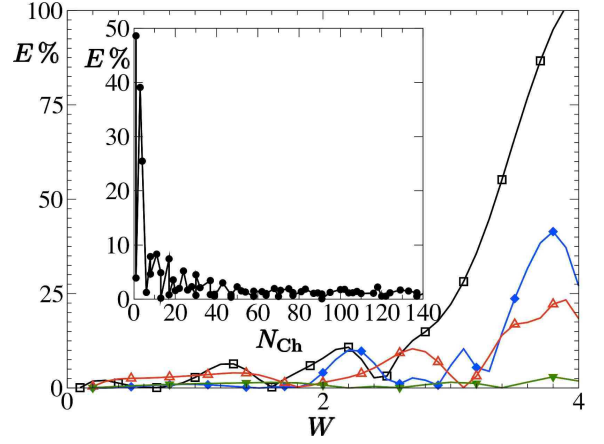}
\caption{Comparison between RMT and TB for a single non magnetic layer. The error  $E\% $   between RMT and TB (see text)
is plotted as a function of the disorder strength $W$ for one single sample.
 The system corresponds to   $L=20$  and  $M=5$   ($N_{\rm ch} =18$, square),
  $M=6$  sites  ($N_{\rm ch} =24$, diamond),  $M=8$  sites  ($N_{\rm ch} =43$, up triangle),   $M=20$  sites  ($N_{\rm ch} =259$, down triangle).
  Inset:~Error  $ E\% $  as a function of the number of open channels $N_{\rm ch}$  for $L=M=20$ with a disorder strength $W=1.5$.}
  \label{fig: figchannels}
\end{figure}
Note that in order to better observe the discrepancy due to fluctuations we perform the calculations on a single sample and do 
{\it not} average over disorder. We checked that different samples give consistent results. 

At first the main panel of Fig.~\ref{fig: figchannels} shows the relative error in function of the disorder strength $W$. 
We observe the effect of  the quantum localization, that  corresponds, as predicted,  to  a clear increase of  the error with the disorder strength. Such rise of  the error can be  tempered  with the increase of  number of  channels, as indicated by the data obtained with $N_{\rm ch}= 259$ (down triangle) . The $E\% $ as a function of $N_{\rm ch}$  is  investigated in the inset of Fig.~\ref{fig: figchannels} ($N_{\rm ch}$ has been varied at fixed $M$ by changing the Fermi energy).
Despite of the strong  fluctuations, this survey  clearly shows  the strong reduction of  the error with the increase of $N_{\rm ch}$.

\begin{figure}
\includegraphics[keepaspectratio,width=0.9\linewidth]{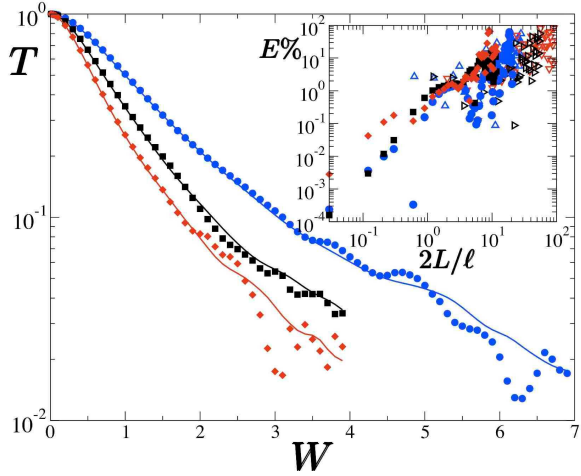}
\caption{
Comparison   between TB (symbols) and RMT  (full line) of the total transmission $T$ as a function of  the disorder strength $W$.
$M=10$ corresponding to  $N_{\rm ch} =66$ open channels and
$L=20$ (circle), $L=40$ (square), $L=60$ (diamond).  
Inset:~Error   $E\% $  in percent between CRMT and TB (see text) in function of  the ratio $L/\ell$.
Full symbol correspond to the data plotted in the main graph. Empty   symbols  correspond to  varying $L$ for fixed $W=2.5$ (up triangle) ,  $W=3.5$ (right triangle) and  $W=4.5$ (down triangle).
We clearly see that we reach the strong localization regime for $L/\ell \geq 10$.
}
\label{fig: figlength} 
\end{figure}

Fig.~\ref{fig: figlength}  shows the total transmission trough the sample as a function of the  disorder strength $W$ for three different  system's size. The data show 
a good agreement between the TB model and CRMT for small disorder. Such agreement  together with   the theoretical prediction of an algebraic decay of  the transmission as a function of  mean free path  $T_{\rm RMT}(L)= (1-  L / \ell )^{-1}$, permit us  to numerically extract the value of   $ \ell = a / W^2$,  with $a=40$ and rescales all the length by  the mean free path the system's size.  The  inset of Fig.~\ref{fig: figlength}  shows  the error in function of the ratio $ 2L/\ell $,  for  various disorder strength  and/or  system's length.  This study indicates that the mixing constraint is  less important than expected, indeed even for  a $ 2L/\ell \leq 0.1 $ the error is almost null. However,  the strong  localization regime is clearly identified for a ratio  $2L/\ell \geq 10 $, for which the error exceeds $50\%$.

\begin{figure}
\includegraphics[keepaspectratio,width=0.9\linewidth]{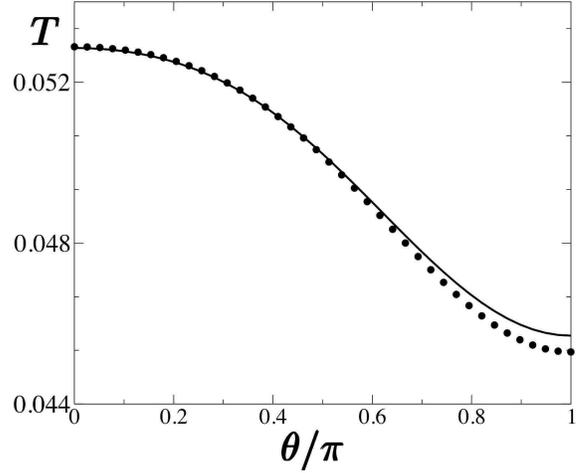}
\caption{\label{fig:multiscale} Comparison   between TB (symbols) and RMT  (full line) of the total transmission $T$  of a  {\bf FIF} multilayer as a function of angle between layers.  
 We note that $L=M=20$ sites that corresponds to $512$ channels,  the disorder strength  is $W^{\u}=0.5$ and $W^{\d}=1.5$. 
 The  insulator is characterized by   $U^{\u}=5$ and $U^{\d}=20$. 
 Data show a good agreement between RMT and TB,  we note the error  $E \%$ is alway bellow one precent.}
\end{figure}

We now turn to the full {\bf FIF}  structure. It is  composed of an  insulator connected to  two ferromagnets. The  first  ferromagnets has a size of $L/2-1$  sites and has a  fixed magnetization, the second one is of size  $3L/2$ with a free magnetization, see  Fig.~\ref{fig:FIF}.   We perform two types of calculations for this sample.  First: we calculate the tunneling magneto-conductance of the whole multilayer using the TB-model, dot point on Fig.~\ref{fig:multiscale}, as a function of the angle  $\theta$ between magnetization of the two {\bf F} layers.  This corresponds to a measure between the probe $0$ and $2$ in Fig.~\ref{fig:FIF}.  Second: we split  the system at the distance L (Probe $1$  in Fig.~\ref{fig:FIF}).  We then construct the corresponding "hat" matrices of each sub-parts and note  $\hat S^{01} $ ($\hat S^{12}$)  for area between probes  $0$ and $1$ ($1$ and $2$).  We then used the "hat" addition law to calculate the "hat" matrices of the full system  ($\hat S^{01}  "+" \hat S^{12}$)  and  plot conductance, solid lines on Fig.~\ref{fig:multiscale}.
Comparison between exact TB-model of the {\bf FIF} layer within RMT show a good agreement, indeed the relative error  $E \%$  for such unique  small sample is  alway less than one percent,  this is quite remarkable and validate the multi-scale approach.    

%

\section{Effective Tight-Binding hamiltonian approach to spin transport} \label{S:TB}

 In this section, we consider  an effective Tight-Binding (TB) Hamiltonian parametrized to reproduce the main bulk and interface properties of a few diffusive magnetic and non magnetic metals. There are several way one can  derive an TB Hamiltonian for magnetic multilayers. For instance, one can use multi-orbital ($s,p,d...$) models where each site of the TB model corresponds to an atom and where all the on site energies and hopping amplitudes are parametrized to reproduce {\it ab-initio} calculations.\cite{Stiles:1996a, Stiles:1996b} When one is interested in transport properties however, one is not interested in a model that reproduces the entire band structure of a material, but only what happens around the Fermi level. Here, we take an even more limited approach: we construct our model such that it correctly reproduces the $5$ characteristic length scales that were introduced in the previous section [mean free path for majority ($l_\u$) and minority ($l_\d$) electrons, spin-flip diffusion length ($l_{\rm sf}$), transverse penetration length ($l_\perp$) and Larmor precession length ($l_{\rm L}$)]. This quantum model can hence be viewed as an effective approach valid for diffusive metals. It is an extension of the toy model of the previous section to properly treat interface properties and spin-flip scattering.

\subsection{A minimum TB model}\label{SS:min_TB}

 We  consider  a rectangular system
of volume $L_x\times L_y\times L_z$  which we discretize on $M_x\times M_y\times M_z$ cells on a cubic lattice, each cell having a small volume $b_x\times b_y\times b_z$. We choose the $x$ axis to be the direction of growth of the pillar.  Each cell corresponds to one site of the TB Hamiltonian which reads, 
\begin{equation}
\label{eq:TB-H}
H=-{\sum_{<i,j>}{\bm c}_{i}^{\dagger}{t}_{ij}{{\bm c}_j}}
+{\sum_i{\bm c}_{i}^{\dagger}{V}_i{{\bm  c}_i}}-{\sum_{<i,j>}{\bm c}_{i}^{\dagger}{t}_{ij}^{so}{{\bm c}_j}}
\end{equation}
The inter-site hopping matrix $t_{ij}$ is material dependent and fix the band structure. We use nearest neighbors hopping with different values for hopping elements along the multilayer growth ($x$)  direction ($t_{\parallel}$) and within the plane ($y,z$) of the layers ($t_{\perp}$). Allowing for different $t_{\parallel}$ and $t_{\perp}$ is advantageous in the simulations: it allows for different discretization steps in the $x$ direction and $y,z$ plane. Also, by taking $t_{\perp}$ significantly smaller than $t_{\parallel}$, one ensures that the number of propagating channels is maximum and constant over a large window of energy, hence reducing mesoscopic effects. On the other hand, the model acquires some intrinsic anisotropy not present in the original materials, so that the  physical meaning of the calculations is doubtful outside the quasi-one dimensional geometry considered here. 
These hopping elements are also spin resolved,
\begin{eqnarray}
\label{eq:t_hopp}
t_{\parallel,\perp} =
\left(
\begin{array}{cc} 
t^{\u}_{\parallel,\perp} & 0 \\ 
0 & t^{\d}_{\parallel,\perp}
\end{array}
\right).
\end{eqnarray}
The second term in the Hamiltonian represents the scattering potential for electrons with different spins  and is still described by the Eq.~(\ref{eq:scat-pot}).
The last term $t^{so}_{ij}$ introduces some spin-orbit interaction responsible for a finite spin-flip scattering length.
Its spin structure respects time-reversal symmetry and is given by,
\begin{equation}\label{eq:t_so}
t_{ij}^{so}=
W_{so} 
\left(
\begin{array}{cc} 
\xi_{ij} & -\eta^{\ast}_{ij}\\
\eta_{ij} & \xi^{\ast}_{ij} 
\end{array}\right),
\end{equation}
where the elements of the matrix are random complex numbers distributed according to the gaussian distribution, with $\overline{\eta^2_{ij}}=\overline{\xi^2_{ij}}=1$. Different pairs of nearest neighbors have independent values of $\eta^2_{ij}$ and $\xi^2_{ij}$. $W_{so}$  represent the strength of spin orbit interaction. The parameters $t^\sigma_{\parallel}$, $t^\sigma_{\perp}$, $W^{\u,\d}$ and $W^{so}$ which define the TB-model allow to tune the different characteristic lengths that have been introduced before.

{\it Interfaces.} In order to complete the model, the above bulk model has to extended to deal with interfaces properties.
Interfaces are represented as hopping elements that link the sites of two different materials. In this view, an interface
acts as a potential barrier between two adjacent materials, and does not contain any disorder or spin orbit interaction.
The hamiltonian that describes interfaces is thus simply given by
\begin{equation}\label{eq:H_int}
H_{int}=-{\sum_{i,j}{\bm c}_{i}^{\dagger}{t}_{ij}^{int}{{\bm c}_j}}
\end{equation} 
where the indexes $i,j$ belong to the sites of the adjacent materials, and the matrix $t^{int}_{ij}$ is given by
\begin{equation}\label{eq:t_int}
t_{ij}^{int}=
\left(
\begin{array}{cc} 
t_{\uparrow} & 0\\
0 & t_{\downarrow} 
\end{array}\right),
\end{equation}

\subsection{Numerical study of a bulk layer within the TB-model.}   %
\begin{figure}
\includegraphics[keepaspectratio,width=0.9\linewidth]{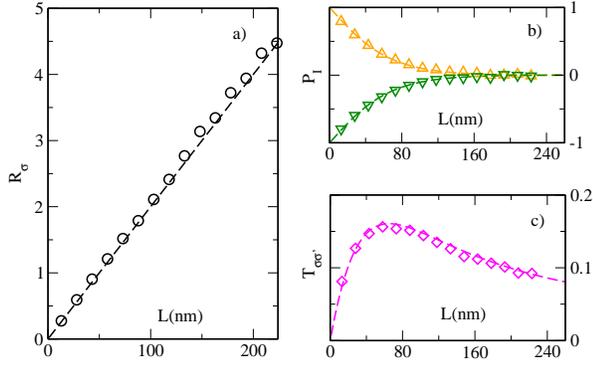}
\caption{\label{fig:gold} Comparison between TB (symbols) and CRMT (dashed lines) calculations for transport properties of a normal metal as a function of the thickness $L$ of the sample. The parameters have been chosen to reproduce the properties of gold (Au). Left panel: intrinsic resistance $R_I^\u(L)=R_I^\d(L)$ in unit of the Sharvin resistance. Right upper panel: $P_I^\u(L)$ (up triangles) and $P_I^\d(L)$ (down triangles). Right lower panel: spin flip transmission probability $T_{\u\d}(L)$. These results have been averaged over 6 samples with different realizations of the disorder.}
\end{figure}

\begin{figure}
\includegraphics[keepaspectratio,width=0.9\linewidth]{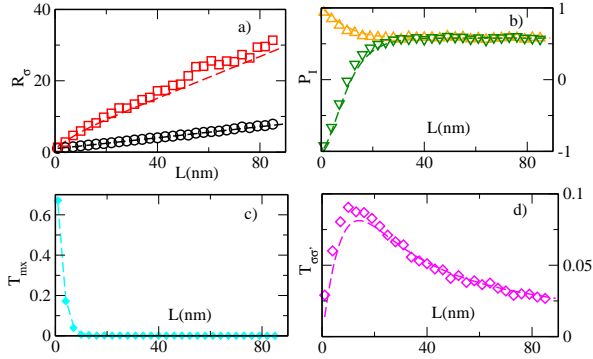}
\caption{\label{fig:permalloy} 
Comparison between TB (symbols) and CRMT (dashed lines) calculations for transport properties of a magnetic metal as a function of the thickness $L$ of the sample. The parameters have been chosen to reproduce the properties of permalloy (Py). Left upper panel: intrinsic resistance $R_I^\u(L)$ (circles) and $R_I^\d(L)$ (squares) in unit of the Sharvin resistance. Left lower: mixing transmission $T_{\rm mx}(L)$. Right upper panel: $P_I^\u(L)$ (up triangles) and $P_I^\d(L)$ (down triangles). Right lower panel: spin flip transmission probability $T_{\u\d}(L)$. These results have been averaged over 6 samples with different realizations of the disorder.}
\end{figure}

\label{SS:TB-params}                                                %
We are now ready to perform a numerical study of our TB model. The quantities of interest are the
intrinsic total resistance $R_I^\sigma$ (in unit $R_{\rm sh}$)  for spin $\sigma$,
\be
R_I^\sigma= \frac{1}{T_{\u\sigma}+T_{\d\sigma}}-1,
\ee
the current polarization $P_I^\sigma$ upon sending polarized electrons along the $\sigma=\u,\d$ 
direction,
\be
P_I^\sigma=\frac{T_{\u\sigma}-T_{\d\sigma}}{T_{\u\sigma}+T_{\d\sigma}},
\ee
the mixing transmission $T_{\rm mx}$ and the spin-flip probability $T_{\u\d}$. Fig.~\ref{fig:gold} and
Fig.~\ref{fig:permalloy} show these quantities for two sets of parameters that correspond respectively
to a normal (Au) and a magnetic (Py) material as a function of the thickness $L$ of the sample.
 Together with the quantum calculations
(symbols) we also plot the results of the CRMT calculations (lines). The latter is equivalent for this collinear case to the Valet Fert Ohmic theory. We find that our TB effective model reproduces extremely well
the semi-classical results: the intrinsic resistance is indeed Ohmic ($R_I^\sigma=\rho^\sigma L$) and
the polarization reaches its asymptotic value $\beta_d$ after an exponential decay controlled by $l_{\rm sf}$.
Exact CRMT expressions for $P_I^\sigma(L)$ are cumbersome but amount with good precision to
\be
P_I^\sigma(L)\approx \beta_d - (\beta_d\pm 1)e^{-L/l_{\rm sf}}
\ee
which is well verified by the TB quantum data. For magnetic material, $T_{\rm mx}$ is also found to decay exponentially, as expected. Such an agreement between the TB model and CRMT allows to use the experimentally measured CRMT (Valet Fert) parameters to tabulate the TB model. The corresponding parameters can be found in Table~\ref{TBbulk}. We note that the polarization saturates towards $\beta_d$ which defers slightly from the Valet-Fert definition of $\beta$ [see Eq.~(\ref{eq:AP:beta_d})] due to the depolarizing role of the Sharvin resistances. This point will be discussed further in Appendix~\ref{App:Direct_C-RMT}.

\begin{table}
\begin{center} 
\begin {tabular}{|c|c|c|c|c|c|c|c|c|c|}
\hline
Material & $\rho*$ & $\beta$ & $1/l_{\rm sf}$ & $W_{\uparrow}$ & $W_\downarrow$ & $W_{so}$ & $t_{\parallel\uparrow}$ & $t_{\parallel\downarrow}$ & $t_\perp$  \\
\hline
Cu & 5 & 0 & 0.002 & 0.45 & 0.45 & 0.0055 & 1 & 1 & 0.4 \\
\hline
Au & 20 & 0 & 0.033 & 0.7 & 0.7 & 0.042 & 1 & 1 & 0.4 \\
\hline
Co & 75 & 0.46 & 0.017 & 4.1 & 2.05 & 0.022 & 3 & 1 & 0.4 \\
\hline
Py & 291 & 0.76 & 0.182 & 3 & 4.6 & 0.11 & 2 & 1 & 0.4 \\
\hline
\end {tabular}
\end{center}
\caption{Bulk TB parameters for a few metals. Valet Fert resistivity $\rho^{\ast}$ is measured in units $10^{-9}\Omega m$ and spin-flip length in $nm$.
The discretization length $b_z$ is equal to $1nm$. These parameters have been obtained averaging the calculation of the resistance over 6 samples with different
realizations of the disorder.  We have used $R_{\rm sh}=2f\Omega .m^2$. }
\label{TBbulk}
\end{table}

\subsection{Interface properties of the TB model} %
\begin{figure}
\includegraphics[keepaspectratio,width=0.8\linewidth]{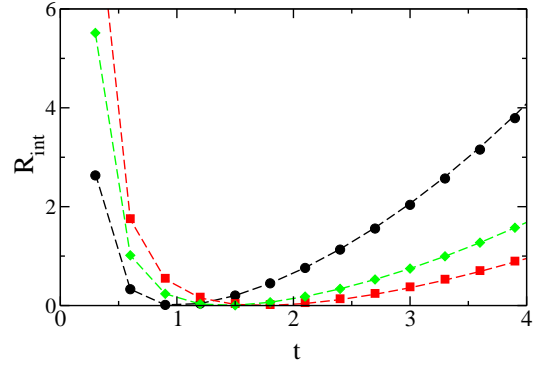}
\caption{\label{fig:interface} Resistance of the interface as a function of the hopping $t$ between two adjacent materials for various 
mismatches of the bulk hopping parameters: 
Symbols represent the interface resistance (for a given spin) between two materials with $t_{\parallel\rm  left}=3$ (red squares),
$t_{\parallel\rm  left}=2$ (green diamonds) and $t_{\parallel\rm left}=1$ (black circles). The hopping parameter of the right material is kept constant $t_{\parallel\rm  right}=1$. The lines are Pad\'e fits of the data of the form $R_{int}=A\times(t^2-t_{\parallel\rm  left}t_{\parallel\rm  right})^2/t^2$ with $A=0.09$, $A=0.14$ and $A=0.29$ respectively.}
\end{figure}

A similar study can be performed for interfaces properties. In this case, the quantity of interest is the interface resistance
$R_{int}^\sigma$ (in unit $R_{\rm sh}$)  for spin $\sigma$,
\begin{equation}
R_{int}^{\sigma}=\frac{1}{T_{\sigma\sigma}}-1
\end{equation}
where $T_{\sigma\sigma}$ represents the transmission probability of the interface. 
The interface resistance has been calculated as a function of the hopping $t_{\uparrow(\downarrow)}$ at the interface, 
and fitted with simple Pad\'e interpolating formulas  (Fig.~{\ref{fig:interface}}).
Knowing how the interface resistance varies as a function of the hopping allows us to use the Valet Fert experimentally
measured parameters to tabulate the TB model for the interfaces, as we have done with the bulk. The TB parameters
calculated for a few materials are shown in Table~\ref{TBint}.

\begin{table}
\begin{center}
\begin {tabular}
{|c|c|c|c|c|}
\hline
Interface & $r_b^{\ast} (10^{-15}\Omega m^2)$ & $\gamma$ & $t_\uparrow$ & $t_\downarrow$ \\
\hline
$Au|Co$ & 0.5 & 0.77 & 1.257 & 2.2 \\
\hline
$Cu|Co$ & 0.51 & 0.77 & 1.253 & 2.21 \\
\hline
$Cu|Py$ & 0.5 & 0.7 & 0.99 & 2.17 \\
\hline
$Au|Py$ & 0.5 & 0.77 & 1.257 & 2.2 \\
\hline
$Au|Cu$ & 0.5 & 0 & 1.86 & 1.86 \\
\hline
\end {tabular}
\end{center}
\caption{TB parameters for a few Normal-Ferromagnetic interfaces. We find parameters of TB model using CRMT parameters for the interfaces Eq.~(\ref{eq:interface}).
The discretization length $b_z$ is equal to $1nm$.}
\label{TBint}
\end{table}

\subsection{Comparison between CRMT and TB for a spin valve.} %
\begin{figure}
\includegraphics[keepaspectratio,width=1.0\linewidth]{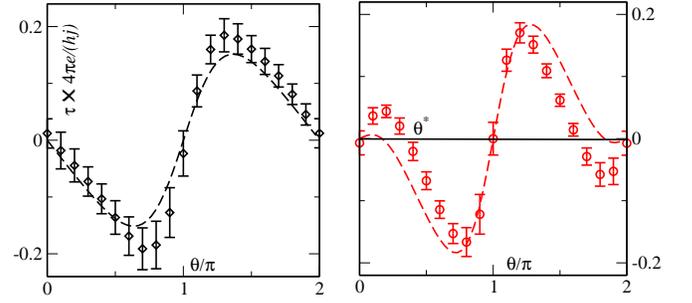}
\caption{\label{fig:ST_wavy_not_wavy} Comparison between TB (symbols) and CRMT (dashed lines) calculations for the angular dependence of the spin torque $\tau$ exerted on the ferromagnet at the right hand side of the stack. Left panel: $Cu_5Py_{20}Cu_5Py_{20}Cu_5$ multilayer (length in $nm$). Right panel: $Cu_{159}Co_{8}Cu_{10}Py_{8}Cu_4$ which shows a wavy behavior.}
\end{figure}

\begin{figure}
\includegraphics[keepaspectratio,width=1.0\linewidth]{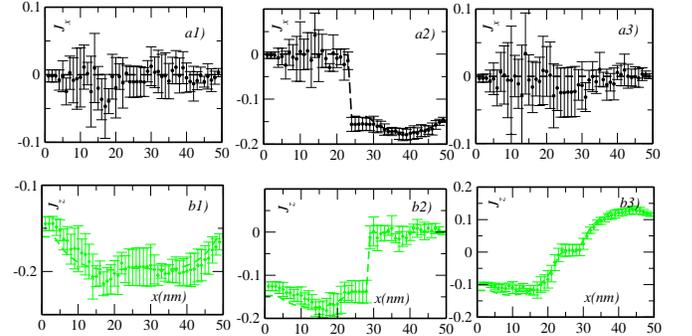}
\caption{\label{fig:CuPyCuPy_angleCu} Comparison between TB (symbols) and C-RMT (dashed lines) calculations for the polarization of the spin current inside the $Cu_5Py_{20}Cu_5Py_{20}(\theta)Cu_5$ stack. Magnetization of the second $Py$ layer is rotated around in $XZ$ plain while the first one point in the $Z$ direction. Left, middle and right panels stand for $\theta=0,\pi/2$ and $\pi$ respectively. Upper (Lower) panels shows the spacial $x$ dependence of the spin current $j_x=J_x/I$ ($j_z=J_z/I$) along  the $X$ ($Z$) direction.}
\end{figure}

Now that our model is fully tabulated and that we have checked that all the individual pieces (bulk and interfaces)
agree with CRMT theory, we can go ahead and perform quantum calculation for entire spin valves. The results
are presented in Fig.~\ref{fig:ST_wavy_not_wavy} where we calculate the spin torque $\tau$ (defined as the difference of spin current on the two sides of the free ferromagnet\cite{Berger:1996, Slonczewski:1996,Waintal:2000}) as a function of the angle $\theta$ between the magnetization of the two magnetic layers. Two different stacks are presented 
(lengths in $nm$): $Cu_5Py_{20}Cu_5Py_{20}Cu_5$ (left panel) and $Cu_{159}Co_{8}Cu_{10}Py_{8}Cu_4$ (right panel). The first one is symmetric and shows a usual $\tau\propto \sin\theta$ torque~\cite{Slonczewski:1996, Berger:1996}. The CRMT and TB calculations are in close agreement. For the second stack which is asymmetric, we find a good agreement of the torques, except at small angle where the TB calculation shows a strong deviation from the $\sin \theta$ behavior. As a result, the TB torque vanishes for a finite value $\theta=\theta^{\ast}$ and the corresponding structure is so called "wavy" (see Ref.~\onlinecite{Manschot:2004, Barnas:2005, Gmitra:2006, Boulle:2007} for an extensive discussion of waviness where neither the parallel nor the antiparallel magnetic configuration is stable). The corresponding CRMT calculation is close, but below,  to the wavy instability threshold~\cite{Fert:2004,Gmitra:2009,Rychkov:2009}. This small discrepancy  indicates a weakness of the effective TB approach as its physics is fairly sensitive to the choice of TB parameters.
To proceed further, we consider in Fig.~\ref{fig:CuPyCuPy_angleCu} the symmetric $Cu_5Py_{20}Cu_5Py_{20}(\theta)Cu_5$ stack and compute the
spin current inside the sample as a function of the position $x$ along the stack. Once again, we observe a
very good agreement between the two approaches. However, this new calculation points again to the main weakness
of the TB approach: the error bars in the TB calculation correspond to the mesoscopic fluctuations upon averaging on different samples (typically 10 in those calculations). Real spin valves nanopillars typically have $N_{\rm ch}\approx 10^4$ channels so that mesoscopic fluctuations $\approx 1/N_{\rm ch}$ are negligible.
Our TB calculations are performed with typically $50-100$ channels, hence do show significant sample to sample
fluctuations so that quantitative correspondence with CRMT calculations are only obtained after averaging over
different disorder configurations.

To conclude the last two sections, we have found a very good agreement between the TB approach and the CRMT approach even in regimes where it was not really expected: the derivation of CRMT assumes that the different channels are ergodically mixed, which is only achieved when the typical thickness of the layers is large compared to their men free path. In that sense, the present TB results can be viewed as a proof of the robustness of the CRMT approach as good agreement is also obtained in fairly transparent regimes. On the other hand, as  TB calculations are numerically much more demanding than CRMT, practical calculations are best performed with the latter approach which do not suffer from mesoscopic fluctuations.

\section{Conclusion}
In conclusion, we have given an extensive presentation of the semi-classical CRMT theory with an emphasis on the different
connections with other existing theories (Generalized circuit theory and Valet-Fert theory). While CRMT is semi-classical, the fact that it has been directly derived from the (microscopic) quantum scattering approach allows for a simple, direct connection to
other quantum mechanical calculations. We have used this connection to study the regime of validity of the semi-classical approximation. In particular we studied its accuracy when part of the system (the insulating layer) is non-Ohmic and must be described at the quantum mechanical level. The results allows to validate a multi-scale approach to magnetic multi-layers where
the quantum calculations provide boundary conditions for the semi-classical simulations. A natural extension of this work is to generalize CRMT to three dimensions, allowing to deal with system with non trivial magnetic textures such as domain wall or vortices\cite{Petitjean:2011} Upon  embedding the resulting theory in micromagnetic simulations, one would reach a level of description that should be able to describe quantitatively a large class of spintronic devices.

   \begin{acknowledgments}
   We thank T. Valet, A. Fert, H. Jaffres, G. Bauer and A. Brataas for valuable discussions. 
   This work was supported by EC Contract No. IST-033749 DynaMax" , CEA NanoSim program,  CEA Eurotalent and EC Contract ICT-257159 ÒMACALOÓ.
    \end{acknowledgments}
\appendix

\section{\label{App:Direct_C-RMT} Direct integration of CRMT equations Eq.~(\ref{drdl}) and Eq.~(\ref{dtdl}).}

In this appendix, we provide a direct integration of the CRMT equations (\ref{drdl}) 	and (\ref{dtdl}) for the bulk properties of a magnetic layer.  An alternative (probably simpler)  way  would be to rely on the mapping with Valet-Fert diffusion equation.

We start with the integration of the "hat" reflection matrix  defined in Eq.~(\ref{drdl}) as it is decoupled from the "hat" transmission matrix  Eq.~(\ref{dtdl}). Let us define the stationary solution of $\frac{\partial\hat r}{\partial L} =0$  to be  $\hat r_0$.   We   seek  solutions of Eq.~(\ref{drdl})   in the  form: $\hat r(L)=\hat r_0 + \hat r_1(L)$ which yields, 
\be
\label{eq:dr1dl}
\frac{\partial\hat r_1}{\partial L} = 
(\hat r_0\Lambda^r-\Lambda^t)\hat r_1 +\hat r_1(\Lambda^r\hat r_0-\Lambda^t) + \hat r_1\Lambda^r\hat r_1
\ee

Since  we consider here system with an homogeneous magnetization, the  inner (mixing) part  and  of  the matrix are  decoupled from the up and down part so that the initial $4\times 4$ matrices can be reduced to  $2\times 2$.  The $2\times 2$ matrix  $r_0$ (reduced version of  $\hat r_0$) satisfies, 
\begin{equation}
\label{eq:r_0}
\tilde\Lambda^r-\tilde\Lambda^tr_0-r_0\tilde\Lambda^t+r_0\tilde\Lambda^rr_0=0
\end{equation}
where the $2\times2$  matrices $\tilde\Lambda^{t/r}$ are defined in Eq.~(\ref{eq:lambda_tilde}).
 After some algebra we  can transform the preceding equation  into
\begin{equation}
\label{eq:eq_for_r_0}
\frac{1-r_0}{1+r_0}\Gamma\frac{1-r_0}{1+r_0}=\Gamma_{\rm sf} Y
\end{equation}
where,  
 \begin{equation}
 \Gamma=
 \begin{pmatrix}
  \Gamma_{\u} & 0 \\
  0 &\Gamma_{\d}
 \end{pmatrix}
 \, \mbox{and } 
 Y=
 \begin{pmatrix}
1 & -1\\
  -1 &1
 \end{pmatrix}
\end{equation}

Such kind of quadratic matrix equations  can be  solved~\cite{Shurbet:1974} and yields
\begin{equation}
\label{eq:solution_r_0}
r_0=\frac{1-\Gamma_{\rm sf}^{1/2}\Gamma^{-1}(\Gamma Y)^{1/2}}{1+\Gamma_{\rm sf}^{1/2}\Gamma^{-1}(\Gamma Y)^{1/2}}
\end{equation}

We proceed with the following change of variable,
\begin{eqnarray}
\label{eq:rbar}
\hat r_1(L) &\equiv&e^{\Omega_BL} \, \bar{r}(L) \,  e^{\Omega_AL}\\
\bar\Lambda(L) &\equiv& e^{\Omega_AL} \Lambda^r  e^{\Omega_BL},
\end{eqnarray}
where $\Omega_A=r_0\Lambda^r-\Lambda^t$ and $\Omega_B=\Lambda^r r_0-\Lambda^t$.  
This transformation simplifies  Eq.~(\ref{eq:dr1dl}) into 
\begin{eqnarray} \label{eq:rbardiff}
\frac{\partial\bar r^{-1}}{\partial L}=-\bar\Lambda(L),
\end{eqnarray}
that can be trivially integrated.   Inserting the solution of Eq.(\ref{eq:rbardiff}) into Eq.(\ref{eq:rbar}), we obtain the solution for the refection "hat" matrix  $\hat r(L)$ . 
The latter  can be then inserted into  Eq.~(\ref{dtdl})  that can be straightforwardly integrated to deliver  $\hat t(L)$  the transmission "hat" matrix  .
The analytical expressions for $\hat r(L)$ and $\hat t(L)$ are rather  cumbersome. In what follows we focus on
the  limiting case of $L\gg l_{\rm sf}$ where compact expressions can be obtained.

\begin{figure}
\includegraphics[keepaspectratio,width=0.9\linewidth]{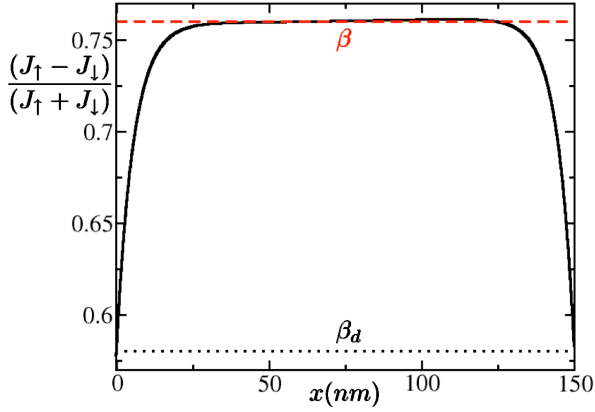}
\caption{\label{fig:Py_inside} Polarization of the spin current inside a Py layer of length 150 $nm$. In the middle of the sample the polarization of the spin current is equal to the bulk  value $\beta$ ($\beta=0.76$, red dashed line) while close to the ends it decreases to the value defined by Eq.~(\ref{eq:spin_polar_sh}), ($\beta_d=0.58$, black doted line).}
\end{figure}

Introducing the resistivity per spin channel $\rho_{\u,\d}$ as the total resistance per spin and per unit length, 
\begin{equation}
\label{eq:AP:rhou}
\frac{\rho_\sigma}{{\cal R}_{\rm sh}} =\frac{1}{ L}\left(\frac{1}{T_{\u\sigma}(L)+T_{\d\sigma}(L)}\right)
\end{equation}
we define the analogs of  the Valet Fert parameters for the resistivity $\rho^{\ast}_d$  and spin asymmetry $\beta_d$  as: 
\begin{eqnarray}
\label{eq:AP:rho*}
\rho^{\ast}_d&=&(\rho_\u+\rho_\d)/4\\
\label{eq:AP:beta}
\beta_d&=&\frac{\rho_\d-\rho_\u}{\rho_\d+\rho_\u}
\end{eqnarray}
Similarly we can  determine the relaxation rate of the polarization of the spin current, (it relaxes as $e^{-L/l^d_{\rm sf}}$) and 
define the corresponding spin-flip length $l^d_{\rm sf}$,
\begin{eqnarray}
\label{eq:AP:lsf}
\frac{1}{l^d_{\rm sf}}&=&-\frac{1}{L}\log\left(\frac{T_{\u\u}(L)-T_{\d\u}(L)}{T_{\u\u}(L)+T_{\d\u}(L)}-\beta_d\right)
\end{eqnarray}

Re-expressing now Eqs.~(\ref{eq:AP:rho*},\ref{eq:AP:beta},\ref{eq:AP:lsf}) with the help of our CRMT parameters defined in Eq.~(\ref{eq:lambda}), 
 we can directly compare the diffusive values $l^d_{\rm sf},\rho^{\ast}_d,\beta_d$ to the one  obtained with VF-CRMT mapping,
  as given  in Eqs.~(\ref{eq:C-RMT-VF-lsf},\ref{eq:C-RMT-VF-beta},\ref{eq:C-RMT-VF-rho*})
 \begin{eqnarray}
\label{eq:AP:lsf_d}
\frac{1}{l^d_{\rm sf}} &=& 2\sqrt{\Gamma_{\rm sf}}\sqrt{\Gamma_\uparrow + \Gamma_\downarrow}=\frac{1}{l_{\rm sf}}\\
\label{eq:AP:beta_d}
\beta_d&=&\frac{\Gamma_\downarrow - \Gamma_\uparrow}{\Gamma_\uparrow + \Gamma_\downarrow+1/l^d_{\rm sf}} \not = \beta \\
\label{eq:AP:rhostar_d}
\frac{\rho^{\ast}_d}{{\cal R}_{\rm sh}}&=&(\Gamma_\uparrow+\Gamma_\downarrow)/4.=\frac{\rho^{\ast}}{{\cal R}_{\rm sh}}
\end{eqnarray}
Remarkably only one parameter differ,  indeed the spin asymmetry  $\beta_d$   contains an additional spin flip length $1/l^d_{\rm sf}$ with respect to $\beta$.
That difference  becomes important only when the average  resistance  within a spin flip length is smaller that the Sharvin resistance  $\rho^{\ast}l_{\rm sf}\ll {\cal R}_{sh}$. This discrepancy  is due  to the implicit presence of the reservoirs in the present calculation which translate (in the VF language) into additional Sharvin resistance in series with the magnetic layer. This additional 
Sharvin resistance induce a depolarization of the current similar to the one observed in VF theory at a Ferromagnetic-Normal
interface where the polarization is reduced to $\beta'$,
\begin{eqnarray}
\label{eq:spin_polar}
\beta'=\frac{\beta}{1+r_{\rm sf}^{N}/(\rho^{\ast}_F l_{\rm sf}) },
\end{eqnarray}
with $r_{\rm sf}^{N}=\rho_N l_{\rm sf}$ being the resistivity of a normal layer within spin flip length.
Here we find in analogy,
\be
\label{eq:spin_polar_sh}
\beta_d=\frac{\beta}{1+{\cal R}_{sh}/(4\rho_{\rm F}^{\ast} l_{\rm sf})}.
\ee
To illustrate this point, Fig.~\ref{fig:Py_inside} shows the polarization of the current inside a Permalloy sample: we find the polarization to $\beta$ far away from the boundary and $\beta_d$ close to the reservoirs. This effect should be kept in mind when comparing quantum results from the Landauer-Buttiker approach with Valet-Fert solutions.



\begin{thebibliography}{83}
\expandafter\ifx\csname natexlab\endcsname\relax\def\natexlab#1{#1}\fi
\expandafter\ifx\csname bibnamefont\endcsname\relax
  \def\bibnamefont#1{#1}\fi
\expandafter\ifx\csname bibfnamefont\endcsname\relax
  \def\bibfnamefont#1{#1}\fi
\expandafter\ifx\csname citenamefont\endcsname\relax
  \def\citenamefont#1{#1}\fi
\expandafter\ifx\csname url\endcsname\relax
  \def\url#1{\texttt{#1}}\fi
\expandafter\ifx\csname urlprefix\endcsname\relax\def\urlprefix{URL }\fi
\providecommand{\bibinfo}[2]{#2}
\providecommand{\eprint}[2][]{\url{#2}}

\bibitem[{\citenamefont{Baibich et~al.}(1988)\citenamefont{Baibich, Broto,
  Fert, Dau, and Petroff}}]{Baibich:1988}
\bibinfo{author}{\bibfnamefont{M.~N.} \bibnamefont{Baibich}},
  \bibinfo{author}{\bibfnamefont{J.~M.} \bibnamefont{Broto}},
  \bibinfo{author}{\bibfnamefont{A.}~\bibnamefont{Fert}},
  \bibinfo{author}{\bibfnamefont{F.~N.~V.} \bibnamefont{Dau}},
  \bibnamefont{and} \bibinfo{author}{\bibfnamefont{F.}~\bibnamefont{Petroff}},
  \bibinfo{journal}{Phys. Rev. Lett.} \textbf{\bibinfo{volume}{61}},
  \bibinfo{pages}{2472} (\bibinfo{year}{1988}).

\bibitem[{\citenamefont{Binasch et~al.}(1989)\citenamefont{Binasch,
  Gr\"{u}nberg, Saurenbach, and Zinn}}]{Binasch:1989}
\bibinfo{author}{\bibfnamefont{G.}~\bibnamefont{Binasch}},
  \bibinfo{author}{\bibfnamefont{P.}~\bibnamefont{Gr\"{u}nberg}},
  \bibinfo{author}{\bibfnamefont{F.}~\bibnamefont{Saurenbach}},
  \bibnamefont{and} \bibinfo{author}{\bibfnamefont{W.}~\bibnamefont{Zinn}},
  \bibinfo{journal}{Phys. Rev. B} \textbf{\bibinfo{volume}{39}},
  \bibinfo{pages}{4828} (\bibinfo{year}{1989}).

\bibitem[{\citenamefont{Julli\`{e}re}(1975)}]{Julliere:1975}
\bibinfo{author}{\bibfnamefont{M.}~\bibnamefont{Julli\`{e}re}},
  \bibinfo{journal}{Phys. Lett. A} \textbf{\bibinfo{volume}{54}},
  \bibinfo{pages}{225} (\bibinfo{year}{1975}).

\bibitem[{\citenamefont{Moodera et~al.}(1995)\citenamefont{Moodera, Kinder,
  Wong, and Meservey}}]{Moodera:1995}
\bibinfo{author}{\bibfnamefont{J.~S.} \bibnamefont{Moodera}},
  \bibinfo{author}{\bibfnamefont{L.~R.} \bibnamefont{Kinder}},
  \bibinfo{author}{\bibfnamefont{T.~M.} \bibnamefont{Wong}}, \bibnamefont{and}
  \bibinfo{author}{\bibfnamefont{R.}~\bibnamefont{Meservey}},
  \bibinfo{journal}{Phys. Rev. Lett.} \textbf{\bibinfo{volume}{74}},
  \bibinfo{pages}{3273} (\bibinfo{year}{1995}).

\bibitem[{\citenamefont{Miyazaki and Tezuka}(1995)}]{Miyazaki:1995}
\bibinfo{author}{\bibfnamefont{T.}~\bibnamefont{Miyazaki}} \bibnamefont{and}
  \bibinfo{author}{\bibfnamefont{N.}~\bibnamefont{Tezuka}},
  \bibinfo{journal}{JMMM} \textbf{\bibinfo{volume}{139}}, \bibinfo{pages}{L231}
  (\bibinfo{year}{1995}).

\bibitem[{\citenamefont{Katine et~al.}(2000)\citenamefont{Katine, Albert,
  Buhrman, Myers, and Ralph}}]{Katine:2000}
\bibinfo{author}{\bibfnamefont{J.}~\bibnamefont{Katine}},
  \bibinfo{author}{\bibfnamefont{F.}~\bibnamefont{Albert}},
  \bibinfo{author}{\bibfnamefont{R.}~\bibnamefont{Buhrman}},
  \bibinfo{author}{\bibfnamefont{E.}~\bibnamefont{Myers}}, \bibnamefont{and}
  \bibinfo{author}{\bibfnamefont{D.}~\bibnamefont{Ralph}},
  \bibinfo{journal}{Phys. Rev. Lett.} \textbf{\bibinfo{volume}{84}},
  \bibinfo{pages}{3149} (\bibinfo{year}{2000}).

\bibitem[{\citenamefont{Fert et~al.}(2004)\citenamefont{Fert, Cros, George,
  Grollier, Jaffres, Hamzic, Vaures, Faini, Youssef, and Gall}}]{Fert:2004}
\bibinfo{author}{\bibfnamefont{A.}~\bibnamefont{Fert}},
  \bibinfo{author}{\bibfnamefont{V.}~\bibnamefont{Cros}},
  \bibinfo{author}{\bibfnamefont{J.-M.} \bibnamefont{George}},
  \bibinfo{author}{\bibfnamefont{J.}~\bibnamefont{Grollier}},
  \bibinfo{author}{\bibfnamefont{H.}~\bibnamefont{Jaffres}},
  \bibinfo{author}{\bibfnamefont{A.}~\bibnamefont{Hamzic}},
  \bibinfo{author}{\bibfnamefont{A.}~\bibnamefont{Vaures}},
  \bibinfo{author}{\bibfnamefont{G.}~\bibnamefont{Faini}},
  \bibinfo{author}{\bibfnamefont{J.~B.} \bibnamefont{Youssef}},
  \bibnamefont{and} \bibinfo{author}{\bibfnamefont{H.~L.} \bibnamefont{Gall}},
  \bibinfo{journal}{JMMM} \textbf{\bibinfo{volume}{272}}, \bibinfo{pages}{1706}
  (\bibinfo{year}{2004}).

\bibitem[{\citenamefont{Devolder et~al.}(2007)\citenamefont{Devolder, Chappert,
  Katine, Carey, and Ito}}]{Devolder:2007}
\bibinfo{author}{\bibfnamefont{T.}~\bibnamefont{Devolder}},
  \bibinfo{author}{\bibfnamefont{C.}~\bibnamefont{Chappert}},
  \bibinfo{author}{\bibfnamefont{J.}~\bibnamefont{Katine}},
  \bibinfo{author}{\bibfnamefont{M.}~\bibnamefont{Carey}}, \bibnamefont{and}
  \bibinfo{author}{\bibfnamefont{K.}~\bibnamefont{Ito}},
  \bibinfo{journal}{Phys. Rev. B} \textbf{\bibinfo{volume}{75}},
  \bibinfo{pages}{064402} (\bibinfo{year}{2007}).

\bibitem[{\citenamefont{Strachan et~al.}(2008)\citenamefont{Strachan,
  Chembrolu, Acremann, Yu, Tulapurkar, Tyliszczak, Katine, Carey, Scheinfein,
  Siegmann et~al.}}]{Strachan:2008}
\bibinfo{author}{\bibfnamefont{J.}~\bibnamefont{Strachan}},
  \bibinfo{author}{\bibfnamefont{V.}~\bibnamefont{Chembrolu}},
  \bibinfo{author}{\bibfnamefont{Y.}~\bibnamefont{Acremann}},
  \bibinfo{author}{\bibfnamefont{X.}~\bibnamefont{Yu}},
  \bibinfo{author}{\bibfnamefont{A.}~\bibnamefont{Tulapurkar}},
  \bibinfo{author}{\bibfnamefont{T.}~\bibnamefont{Tyliszczak}},
  \bibinfo{author}{\bibfnamefont{J.}~\bibnamefont{Katine}},
  \bibinfo{author}{\bibfnamefont{M.}~\bibnamefont{Carey}},
  \bibinfo{author}{\bibfnamefont{M.}~\bibnamefont{Scheinfein}},
  \bibinfo{author}{\bibfnamefont{H.}~\bibnamefont{Siegmann}},
  \bibnamefont{et~al.}, \bibinfo{journal}{Phys. Rev. Lett.}
  \textbf{\bibinfo{volume}{100}}, \bibinfo{pages}{247201}
  (\bibinfo{year}{2008}).

\bibitem[{\citenamefont{Kiselev et~al.}(2003)\citenamefont{Kiselev, Sankey,
  Krivorotov, Emley, Schoelkopf, Buhrman, and Ralph}}]{Kiselev:2003}
\bibinfo{author}{\bibfnamefont{S.~I.} \bibnamefont{Kiselev}},
  \bibinfo{author}{\bibfnamefont{J.~C.} \bibnamefont{Sankey}},
  \bibinfo{author}{\bibfnamefont{I.~N.} \bibnamefont{Krivorotov}},
  \bibinfo{author}{\bibfnamefont{N.~C.} \bibnamefont{Emley}},
  \bibinfo{author}{\bibfnamefont{R.~J.} \bibnamefont{Schoelkopf}},
  \bibinfo{author}{\bibfnamefont{R.~A.} \bibnamefont{Buhrman}},
  \bibnamefont{and} \bibinfo{author}{\bibfnamefont{D.~C.} \bibnamefont{Ralph}},
  \bibinfo{journal}{Nature} \textbf{\bibinfo{volume}{425}},
  \bibinfo{pages}{380} (\bibinfo{year}{2003}).

\bibitem[{\citenamefont{Pufall et~al.}(2006)\citenamefont{Pufall, Rippard,
  Russek, Kaka, and Katine}}]{Pufall:2006}
\bibinfo{author}{\bibfnamefont{M.}~\bibnamefont{Pufall}},
  \bibinfo{author}{\bibfnamefont{W.}~\bibnamefont{Rippard}},
  \bibinfo{author}{\bibfnamefont{S.}~\bibnamefont{Russek}},
  \bibinfo{author}{\bibfnamefont{S.}~\bibnamefont{Kaka}}, \bibnamefont{and}
  \bibinfo{author}{\bibfnamefont{J.}~\bibnamefont{Katine}},
  \bibinfo{journal}{Phys. Rev. Lett.} \textbf{\bibinfo{volume}{97}},
  \bibinfo{pages}{087206} (\bibinfo{year}{2006}).

\bibitem[{\citenamefont{Boone et~al.}(2009)\citenamefont{Boone, Katine,
  Childress, Zhu, Cheng, and Krivorotov}}]{Boone:2009}
\bibinfo{author}{\bibfnamefont{C.}~\bibnamefont{Boone}},
  \bibinfo{author}{\bibfnamefont{J.}~\bibnamefont{Katine}},
  \bibinfo{author}{\bibfnamefont{J.}~\bibnamefont{Childress}},
  \bibinfo{author}{\bibfnamefont{J.}~\bibnamefont{Zhu}},
  \bibinfo{author}{\bibfnamefont{X.}~\bibnamefont{Cheng}}, \bibnamefont{and}
  \bibinfo{author}{\bibfnamefont{I.}~\bibnamefont{Krivorotov}},
  \bibinfo{journal}{Phys. Rev. B} \textbf{\bibinfo{volume}{79}},
  \bibinfo{pages}{140404} (\bibinfo{year}{2009}).

\bibitem[{\citenamefont{Houssameddine et~al.}(2009)\citenamefont{Houssameddine,
  Ebels, Dieny, Garello, Michel, Delaet, Viala, Cyrille, Katine, and
  Mauri}}]{Houssameddine:2009}
\bibinfo{author}{\bibfnamefont{D.}~\bibnamefont{Houssameddine}},
  \bibinfo{author}{\bibfnamefont{U.}~\bibnamefont{Ebels}},
  \bibinfo{author}{\bibfnamefont{B.}~\bibnamefont{Dieny}},
  \bibinfo{author}{\bibfnamefont{K.}~\bibnamefont{Garello}},
  \bibinfo{author}{\bibfnamefont{J.-P.} \bibnamefont{Michel}},
  \bibinfo{author}{\bibfnamefont{B.}~\bibnamefont{Delaet}},
  \bibinfo{author}{\bibfnamefont{B.}~\bibnamefont{Viala}},
  \bibinfo{author}{\bibfnamefont{M.-C.} \bibnamefont{Cyrille}},
  \bibinfo{author}{\bibfnamefont{J.}~\bibnamefont{Katine}}, \bibnamefont{and}
  \bibinfo{author}{\bibfnamefont{D.}~\bibnamefont{Mauri}},
  \bibinfo{journal}{Phys. Rev. Lett.} \textbf{\bibinfo{volume}{102}},
  \bibinfo{pages}{257202} (\bibinfo{year}{2009}).

\bibitem[{\citenamefont{Villard et~al.}(2010)\citenamefont{Villard, Ebels,
  Houssameddine, Katine, Mauri, Delaet, Vincent, Cyrille, Viala, Michel
  et~al.}}]{Villard:2010}
\bibinfo{author}{\bibfnamefont{P.}~\bibnamefont{Villard}},
  \bibinfo{author}{\bibfnamefont{U.}~\bibnamefont{Ebels}},
  \bibinfo{author}{\bibfnamefont{D.}~\bibnamefont{Houssameddine}},
  \bibinfo{author}{\bibfnamefont{J.}~\bibnamefont{Katine}},
  \bibinfo{author}{\bibfnamefont{D.}~\bibnamefont{Mauri}},
  \bibinfo{author}{\bibfnamefont{B.}~\bibnamefont{Delaet}},
  \bibinfo{author}{\bibfnamefont{P.}~\bibnamefont{Vincent}},
  \bibinfo{author}{\bibfnamefont{M.-C.} \bibnamefont{Cyrille}},
  \bibinfo{author}{\bibfnamefont{B.}~\bibnamefont{Viala}},
  \bibinfo{author}{\bibfnamefont{J.-P.} \bibnamefont{Michel}},
  \bibnamefont{et~al.}, \bibinfo{journal}{IEEE Jour. Solide-State Circuits}
  \textbf{\bibinfo{volume}{45}}, \bibinfo{pages}{214} (\bibinfo{year}{2010}).
  
\bibitem[{\citenamefont{Chappert et~al.}(2007)\citenamefont{Chappert, Fert, and
  Dau}}]{Chappert:2007}
\bibinfo{author}{\bibfnamefont{C.}~\bibnamefont{Chappert}},
  \bibinfo{author}{\bibfnamefont{A.}~\bibnamefont{Fert}}, \bibnamefont{and}
  \bibinfo{author}{\bibfnamefont{F.}~\bibnamefont{Dau}},
  \bibinfo{journal}{Nature Mat.} \textbf{\bibinfo{volume}{6}},
  \bibinfo{pages}{813} (\bibinfo{year}{2007}).

\bibitem[{\citenamefont{Braun et~al.}(2004)\citenamefont{Braun, K\"{o}nig, and
  Martinek}}]{Braun:2004}
\bibinfo{author}{\bibfnamefont{M.}~\bibnamefont{Braun}},
  \bibinfo{author}{\bibfnamefont{J.}~\bibnamefont{K\"{o}nig}},
  \bibnamefont{and} \bibinfo{author}{\bibfnamefont{J.}~\bibnamefont{Martinek}},
  \bibinfo{journal}{Phys. Rev. B} \textbf{\bibinfo{volume}{70}},
  \bibinfo{pages}{195345} (\bibinfo{year}{2004}).

\bibitem[{\citenamefont{Duine et~al.}(2007)\citenamefont{Duine, nez, Sinova,
  and Macdonald}}]{Duine:2007}
\bibinfo{author}{\bibfnamefont{R.}~\bibnamefont{Duine}},
  \bibinfo{author}{\bibfnamefont{A.~N.} \bibnamefont{nez}},
  \bibinfo{author}{\bibfnamefont{J.}~\bibnamefont{Sinova}}, \bibnamefont{and}
  \bibinfo{author}{\bibfnamefont{A.}~\bibnamefont{Macdonald}},
  \bibinfo{journal}{Phys. Rev. B} \textbf{\bibinfo{volume}{75}},
  \bibinfo{pages}{214420} (\bibinfo{year}{2007}).

\bibitem[{\citenamefont{Stiles and
  Zangwill}(2002{\natexlab{a}})}]{Stiles:2002a}
\bibinfo{author}{\bibfnamefont{M.~D.} \bibnamefont{Stiles}} \bibnamefont{and}
  \bibinfo{author}{\bibfnamefont{A.}~\bibnamefont{Zangwill}},
  \bibinfo{journal}{J. Appl. Phys.} \textbf{\bibinfo{volume}{91}},
  \bibinfo{pages}{6812} (\bibinfo{year}{2002}{\natexlab{a}}).

\bibitem[{\citenamefont{Xiao et~al.}(2004)\citenamefont{Xiao, Zangwill, and
  Stiles}}]{Xiao:2004}
\bibinfo{author}{\bibfnamefont{J.}~\bibnamefont{Xiao}},
  \bibinfo{author}{\bibfnamefont{A.}~\bibnamefont{Zangwill}}, \bibnamefont{and}
  \bibinfo{author}{\bibfnamefont{M.}~\bibnamefont{Stiles}},
  \bibinfo{journal}{Phys. Rev. B} \textbf{\bibinfo{volume}{70}},
  \bibinfo{pages}{172405} (\bibinfo{year}{2004}).

\bibitem[{\citenamefont{Xiao et~al.}(2007)\citenamefont{Xiao, Zangwill, and
  Stiles}}]{Xiao:2007}
\bibinfo{author}{\bibfnamefont{J.}~\bibnamefont{Xiao}},
  \bibinfo{author}{\bibfnamefont{A.}~\bibnamefont{Zangwill}}, \bibnamefont{and}
  \bibinfo{author}{\bibfnamefont{M.}~\bibnamefont{Stiles}},
  \bibinfo{journal}{Eur. Phys. J. B} \textbf{\bibinfo{volume}{59}},
  \bibinfo{pages}{415} (\bibinfo{year}{2007}).

\bibitem[{\citenamefont{Brataas et~al.}(2000)\citenamefont{Brataas, Nazarov,
  and Bauer}}]{Brataas:2000}
\bibinfo{author}{\bibfnamefont{A.}~\bibnamefont{Brataas}},
  \bibinfo{author}{\bibfnamefont{Y.~V.} \bibnamefont{Nazarov}},
  \bibnamefont{and} \bibinfo{author}{\bibfnamefont{G.~E.~W.}
  \bibnamefont{Bauer}}, \bibinfo{journal}{Phys. Rev. Lett.}
  \textbf{\bibinfo{volume}{84}}, \bibinfo{pages}{2481} (\bibinfo{year}{2000}).

\bibitem[{\citenamefont{Brataas et~al.}(2001)\citenamefont{Brataas, Nazarov,
  and Bauer}}]{Brataas:2001}
\bibinfo{author}{\bibfnamefont{A.}~\bibnamefont{Brataas}},
  \bibinfo{author}{\bibfnamefont{Y.}~\bibnamefont{Nazarov}}, \bibnamefont{and}
  \bibinfo{author}{\bibfnamefont{G.~E.~W.} \bibnamefont{Bauer}},
  \bibinfo{journal}{Eur. Phys. J. B} \textbf{\bibinfo{volume}{22}},
  \bibinfo{pages}{99} (\bibinfo{year}{2001}).

\bibitem[{\citenamefont{Bauer et~al.}(2003{\natexlab{a}})\citenamefont{Bauer,
  Tserkovnyak, Huertas-Hernando, and Brataas}}]{Bauer:2003a}
\bibinfo{author}{\bibfnamefont{G.}~\bibnamefont{Bauer}},
  \bibinfo{author}{\bibfnamefont{Y.}~\bibnamefont{Tserkovnyak}},
  \bibinfo{author}{\bibfnamefont{D.}~\bibnamefont{Huertas-Hernando}},
  \bibnamefont{and} \bibinfo{author}{\bibfnamefont{A.}~\bibnamefont{Brataas}},
  \bibinfo{journal}{Adv. in Solid State Physics} \textbf{\bibinfo{volume}{43}},
  \bibinfo{pages}{92} (\bibinfo{year}{2003}{\natexlab{a}}).

\bibitem[{\citenamefont{Bauer et~al.}(2003{\natexlab{b}})\citenamefont{Bauer,
  Tserkovnyak, Huertas-Hernando, and Brataas}}]{Bauer:2003b}
\bibinfo{author}{\bibfnamefont{G.~E.~W.} \bibnamefont{Bauer}},
  \bibinfo{author}{\bibfnamefont{Y.}~\bibnamefont{Tserkovnyak}},
  \bibinfo{author}{\bibfnamefont{D.}~\bibnamefont{Huertas-Hernando}},
  \bibnamefont{and} \bibinfo{author}{\bibfnamefont{A.}~\bibnamefont{Brataas}},
  \bibinfo{journal}{Phys. Rev. B} \textbf{\bibinfo{volume}{67}},
  \bibinfo{pages}{094421} (\bibinfo{year}{2003}{\natexlab{b}}).

\bibitem[{\citenamefont{Tserkovnyak et~al.}(2005)\citenamefont{Tserkovnyak,
  Brataas, Bauer, and Halperin}}]{Tserkovnyak:2005}
\bibinfo{author}{\bibfnamefont{Y.}~\bibnamefont{Tserkovnyak}},
  \bibinfo{author}{\bibfnamefont{A.}~\bibnamefont{Brataas}},
  \bibinfo{author}{\bibfnamefont{G.~E.~W.} \bibnamefont{Bauer}},
  \bibnamefont{and} \bibinfo{author}{\bibfnamefont{B.~I.}
  \bibnamefont{Halperin}}, \bibinfo{journal}{Rev. Mod. Phys.}
  \textbf{\bibinfo{volume}{77}}, \bibinfo{pages}{1375} (\bibinfo{year}{2005}).

\bibitem[{\citenamefont{Brataas et~al.}(2006)\citenamefont{Brataas, Bauer, and
  Kelly}}]{Brataas:2006}
\bibinfo{author}{\bibfnamefont{A.}~\bibnamefont{Brataas}},
  \bibinfo{author}{\bibfnamefont{G.~E.~W.} \bibnamefont{Bauer}},
  \bibnamefont{and} \bibinfo{author}{\bibfnamefont{P.}~\bibnamefont{Kelly}},
  \bibinfo{journal}{Phys. Rep.} \textbf{\bibinfo{volume}{427}},
  \bibinfo{pages}{157} (\bibinfo{year}{2006}).

\bibitem[{\citenamefont{Manchon and Slonczewski}(2006)}]{Manchon:2006}
\bibinfo{author}{\bibfnamefont{A.}~\bibnamefont{Manchon}} \bibnamefont{and}
  \bibinfo{author}{\bibfnamefont{J.}~\bibnamefont{Slonczewski}},
  \bibinfo{journal}{Phys. Rev. B} \textbf{\bibinfo{volume}{73}},
  \bibinfo{pages}{184419} (\bibinfo{year}{2006}).

\bibitem[{\citenamefont{Waintal et~al.}(2000)\citenamefont{Waintal, Myers,
  Brouwer, and Ralph}}]{Waintal:2000}
\bibinfo{author}{\bibfnamefont{X.}~\bibnamefont{Waintal}},
  \bibinfo{author}{\bibfnamefont{E.~B.} \bibnamefont{Myers}},
  \bibinfo{author}{\bibfnamefont{P.~W.} \bibnamefont{Brouwer}},
  \bibnamefont{and} \bibinfo{author}{\bibfnamefont{D.~C.} \bibnamefont{Ralph}},
  \bibinfo{journal}{Phys. Rev. B} \textbf{\bibinfo{volume}{62}},
  \bibinfo{pages}{12317} (\bibinfo{year}{2000}).

\bibitem[{\citenamefont{Rychkov et~al.}(2009)\citenamefont{Rychkov, Borlenghi,
  Jaffres, Fert, and Waintal}}]{Rychkov:2009}
\bibinfo{author}{\bibfnamefont{V.~S.} \bibnamefont{Rychkov}},
  \bibinfo{author}{\bibfnamefont{S.}~\bibnamefont{Borlenghi}},
  \bibinfo{author}{\bibfnamefont{H.}~\bibnamefont{Jaffres}},
  \bibinfo{author}{\bibfnamefont{A.}~\bibnamefont{Fert}}, \bibnamefont{and}
  \bibinfo{author}{\bibfnamefont{X.}~\bibnamefont{Waintal}},
  \bibinfo{journal}{Phys. Rev. Lett.} \textbf{\bibinfo{volume}{103}},
  \bibinfo{pages}{066602} (\bibinfo{year}{2009}).

\bibitem[{\citenamefont{Haney et~al.}(2007{\natexlab{a}})\citenamefont{Haney,
  Waldron, Duine, {n}ez, Guo, and Macdonald}}]{Haney:2007a}
\bibinfo{author}{\bibfnamefont{P.}~\bibnamefont{Haney}},
  \bibinfo{author}{\bibfnamefont{D.}~\bibnamefont{Waldron}},
  \bibinfo{author}{\bibfnamefont{R.}~\bibnamefont{Duine}},
  \bibinfo{author}{\bibfnamefont{A.~N.} \bibnamefont{{n}ez}},
  \bibinfo{author}{\bibfnamefont{H.}~\bibnamefont{Guo}}, \bibnamefont{and}
  \bibinfo{author}{\bibfnamefont{A.}~\bibnamefont{Macdonald}},
  \bibinfo{journal}{Phys. Rev. B} \textbf{\bibinfo{volume}{75}},
  \bibinfo{pages}{174428} (\bibinfo{year}{2007}{\natexlab{a}}).

\bibitem[{\citenamefont{Haney et~al.}(2007{\natexlab{b}})\citenamefont{Haney,
  Waldron, Duine, {n}ez, Guo, and MacDonald}}]{Haney:2007b}
\bibinfo{author}{\bibfnamefont{P.~M.} \bibnamefont{Haney}},
  \bibinfo{author}{\bibfnamefont{D.}~\bibnamefont{Waldron}},
  \bibinfo{author}{\bibfnamefont{R.~A.} \bibnamefont{Duine}},
  \bibinfo{author}{\bibfnamefont{A.~S.~N.} \bibnamefont{{n}ez}},
  \bibinfo{author}{\bibfnamefont{H.}~\bibnamefont{Guo}}, \bibnamefont{and}
  \bibinfo{author}{\bibfnamefont{A.~H.} \bibnamefont{MacDonald}},
  \bibinfo{journal}{Phys. Rev. B} \textbf{\bibinfo{volume}{76}},
  \bibinfo{pages}{024404} (\bibinfo{year}{2007}{\natexlab{b}}).

\bibitem[{\citenamefont{Waldron et~al.}(2007)\citenamefont{Waldron, Liu, and
  Guo}}]{Waldron:2007}
\bibinfo{author}{\bibfnamefont{D.}~\bibnamefont{Waldron}},
  \bibinfo{author}{\bibfnamefont{L.}~\bibnamefont{Liu}}, \bibnamefont{and}
  \bibinfo{author}{\bibfnamefont{H.}~\bibnamefont{Guo}},
  \bibinfo{journal}{Nanotech.} \textbf{\bibinfo{volume}{18}},
  \bibinfo{pages}{424026} (\bibinfo{year}{2007}).

\bibitem[{\citenamefont{Xu et~al.}(2008)\citenamefont{Xu, Wang, and
  Xia}}]{Xu:2008}
\bibinfo{author}{\bibfnamefont{Y.}~\bibnamefont{Xu}},
  \bibinfo{author}{\bibfnamefont{S.}~\bibnamefont{Wang}}, \bibnamefont{and}
  \bibinfo{author}{\bibfnamefont{K.}~\bibnamefont{Xia}},
  \bibinfo{journal}{Phys. Rev. Lett.} \textbf{\bibinfo{volume}{100}},
  \bibinfo{pages}{226602} (\bibinfo{year}{2008}).

\bibitem[{\citenamefont{Fidler and Schrefl}(2000)}]{Fidler:2000}
\bibinfo{author}{\bibfnamefont{J.}~\bibnamefont{Fidler}} \bibnamefont{and}
  \bibinfo{author}{\bibfnamefont{T.}~\bibnamefont{Schrefl}},
  \bibinfo{journal}{J. Phys. D} \textbf{\bibinfo{volume}{33}},
  \bibinfo{pages}{R135} (\bibinfo{year}{2000}).

\bibitem[{\citenamefont{Lee et~al.}(2004)\citenamefont{Lee, Deac, Redon,
  Nozi\`eres, and Dieny}}]{Lee:2004}
\bibinfo{author}{\bibfnamefont{K.-J.} \bibnamefont{Lee}},
  \bibinfo{author}{\bibfnamefont{A.}~\bibnamefont{Deac}},
  \bibinfo{author}{\bibfnamefont{O.}~\bibnamefont{Redon}},
  \bibinfo{author}{\bibfnamefont{J.-P.} \bibnamefont{Nozi\`eres}},
  \bibnamefont{and} \bibinfo{author}{\bibfnamefont{B.}~\bibnamefont{Dieny}},
  \bibinfo{journal}{Nature Mat.} \textbf{\bibinfo{volume}{3}},
  \bibinfo{pages}{877} (\bibinfo{year}{2004}).

\bibitem[{\citenamefont{Xiao et~al.}(2005)\citenamefont{Xiao, Zangwill, and
  Stiles}}]{Xiao:2005}
\bibinfo{author}{\bibfnamefont{J.}~\bibnamefont{Xiao}},
  \bibinfo{author}{\bibfnamefont{A.}~\bibnamefont{Zangwill}}, \bibnamefont{and}
  \bibinfo{author}{\bibfnamefont{M.~D.} \bibnamefont{Stiles}},
  \bibinfo{journal}{Phys. Rev. B} \textbf{\bibinfo{volume}{72}},
  \bibinfo{pages}{014446} (\bibinfo{year}{2005}).

\bibitem[{\citenamefont{Fischbacher et~al.}(2007)\citenamefont{Fischbacher,
  Franchin, Bordignon, and Fangohr}}]{Fischbacher:2007}
\bibinfo{author}{\bibfnamefont{T.}~\bibnamefont{Fischbacher}},
  \bibinfo{author}{\bibfnamefont{M.}~\bibnamefont{Franchin}},
  \bibinfo{author}{\bibfnamefont{G.}~\bibnamefont{Bordignon}},
  \bibnamefont{and} \bibinfo{author}{\bibfnamefont{H.}~\bibnamefont{Fangohr}},
  \bibinfo{journal}{Mag IEEE} \textbf{\bibinfo{volume}{43}},
  \bibinfo{pages}{2896} (\bibinfo{year}{2007}).

\bibitem[{\citenamefont{Berkov and Milta}(2008)}]{Berkov:2008}
\bibinfo{author}{\bibfnamefont{D.}~\bibnamefont{Berkov}} \bibnamefont{and}
  \bibinfo{author}{\bibfnamefont{J.}~\bibnamefont{Milta}},
  \bibinfo{journal}{JMMM} \textbf{\bibinfo{volume}{320}}, \bibinfo{pages}{1238}
  (\bibinfo{year}{2008}).

\bibitem[{\citenamefont{Chen et~al.}(2009)\citenamefont{Chen, Loubens,
  Beaujour, Sun, and Kent}}]{Chen:2009}
\bibinfo{author}{\bibfnamefont{W.}~\bibnamefont{Chen}},
  \bibinfo{author}{\bibfnamefont{G.~D.} \bibnamefont{Loubens}},
  \bibinfo{author}{\bibfnamefont{J.-M.~L.} \bibnamefont{Beaujour}},
  \bibinfo{author}{\bibfnamefont{J.~Z.} \bibnamefont{Sun}}, \bibnamefont{and}
  \bibinfo{author}{\bibfnamefont{A.~D.} \bibnamefont{Kent}},
  \bibinfo{journal}{Appl. Phys. Lett} \textbf{\bibinfo{volume}{95}},
  \bibinfo{pages}{172513} (\bibinfo{year}{2009}).

\bibitem[{\citenamefont{Valet and Fert}(1993)}]{Valet:1993}
\bibinfo{author}{\bibfnamefont{T.}~\bibnamefont{Valet}} \bibnamefont{and}
  \bibinfo{author}{\bibfnamefont{A.}~\bibnamefont{Fert}},
  \bibinfo{journal}{Phys. Rev. B} \textbf{\bibinfo{volume}{48}},
  \bibinfo{pages}{7099} (\bibinfo{year}{1993}).

\bibitem[{\citenamefont{B\"uttiker}(1986)}]{Buttiker:1986}
\bibinfo{author}{\bibfnamefont{M.}~\bibnamefont{B\"uttiker}},
  \bibinfo{journal}{Phys. Rev. B} \textbf{\bibinfo{volume}{33}},
  \bibinfo{pages}{3020} (\bibinfo{year}{1986}).

\bibitem[{\citenamefont{Brouwer and Beenakker}(1995)}]{Brouwer:1995}
\bibinfo{author}{\bibfnamefont{P.~W.} \bibnamefont{Brouwer}} \bibnamefont{and}
  \bibinfo{author}{\bibfnamefont{C.~W.~J.} \bibnamefont{Beenakker}},
  \bibinfo{journal}{Phys. Rev. B} \textbf{\bibinfo{volume}{51}},
  \bibinfo{pages}{7739} (\bibinfo{year}{1995}).

\bibitem[{\citenamefont{Brouwer and Beenakker}(1997)}]{Brouwer:1997}
\bibinfo{author}{\bibfnamefont{P.~W.} \bibnamefont{Brouwer}} \bibnamefont{and}
  \bibinfo{author}{\bibfnamefont{C.~W.~J.} \bibnamefont{Beenakker}},
  \bibinfo{journal}{Phys. Rev. B} \textbf{\bibinfo{volume}{55}},
  \bibinfo{pages}{4695} (\bibinfo{year}{1997}).

\bibitem[{\citenamefont{Beenakker}(1997)}]{Beenakker:1997}
\bibinfo{author}{\bibfnamefont{C.~W.~J.} \bibnamefont{Beenakker}},
  \bibinfo{journal}{Rev. Mod. Phys.} \textbf{\bibinfo{volume}{69}},
  \bibinfo{pages}{731} (\bibinfo{year}{1997}).

\bibitem[{\citenamefont{Landauer}(1957)}]{Landauer:1957}
\bibinfo{author}{\bibfnamefont{D.}~\bibnamefont{Landauer}},
  \bibinfo{journal}{IBM J. Res. Dev.} \textbf{\bibinfo{volume}{1}},
  \bibinfo{pages}{223} (\bibinfo{year}{1957}).

\bibitem[{\citenamefont{Landauer}(1970)}]{Landauer:1970}
\bibinfo{author}{\bibfnamefont{R.}~\bibnamefont{Landauer}},
  \bibinfo{journal}{Philos Mag.} \textbf{\bibinfo{volume}{21}},
  \bibinfo{pages}{863} (\bibinfo{year}{1970}).

\bibitem[{\citenamefont{B\"{u}ttiker et~al.}(1985)\citenamefont{B\"{u}ttiker,
  Imry, Landauer, and Pinhas}}]{Buttiker:1985}
\bibinfo{author}{\bibfnamefont{M.}~\bibnamefont{B\"{u}ttiker}},
  \bibinfo{author}{\bibfnamefont{Y.}~\bibnamefont{Imry}},
  \bibinfo{author}{\bibfnamefont{R.}~\bibnamefont{Landauer}}, \bibnamefont{and}
  \bibinfo{author}{\bibfnamefont{S.}~\bibnamefont{Pinhas}},
  \bibinfo{journal}{Phys. Rev. B} \textbf{\bibinfo{volume}{31}},
  \bibinfo{pages}{6207} (\bibinfo{year}{1985}).

\bibitem[{\citenamefont{Bruno and Chappert}(1992)}]{Bruno:1992}
\bibinfo{author}{\bibfnamefont{P.}~\bibnamefont{Bruno}} \bibnamefont{and}
  \bibinfo{author}{\bibfnamefont{C.}~\bibnamefont{Chappert}},
  \bibinfo{journal}{Phys. Rev. B} \textbf{\bibinfo{volume}{46}},
  \bibinfo{pages}{261} (\bibinfo{year}{1992}).

\bibitem[{\citenamefont{Bruno}(1995)}]{Bruno:1995}
\bibinfo{author}{\bibfnamefont{P.}~\bibnamefont{Bruno}},
  \bibinfo{journal}{Phys. Rev. B} \textbf{\bibinfo{volume}{52}},
  \bibinfo{pages}{411} (\bibinfo{year}{1995}).

\bibitem[{\citenamefont{Waintal and Brouwer}(2002)}]{Waintal:2002}
\bibinfo{author}{\bibfnamefont{X.}~\bibnamefont{Waintal}} \bibnamefont{and}
  \bibinfo{author}{\bibfnamefont{P.~W.} \bibnamefont{Brouwer}},
  \bibinfo{journal}{Phys. Rev. B} \textbf{\bibinfo{volume}{65}},
   \bibinfo{pages}{054407}
  (\bibinfo{year}{2002}).

\bibitem[{\citenamefont{Heinrich et~al.}(2003)\citenamefont{Heinrich,
  Tserkovnyak, Woltersdorf, Brataas, Urban, and Bauer}}]{Heinrich:2003}
\bibinfo{author}{\bibfnamefont{B.}~\bibnamefont{Heinrich}},
  \bibinfo{author}{\bibfnamefont{Y.}~\bibnamefont{Tserkovnyak}},
  \bibinfo{author}{\bibfnamefont{G.}~\bibnamefont{Woltersdorf}},
  \bibinfo{author}{\bibfnamefont{A.}~\bibnamefont{Brataas}},
  \bibinfo{author}{\bibfnamefont{R.}~\bibnamefont{Urban}}, \bibnamefont{and}
  \bibinfo{author}{\bibfnamefont{G.~E.~W.} \bibnamefont{Bauer}},
  \bibinfo{journal}{Phys. Rev. Lett.} \textbf{\bibinfo{volume}{90}},
  \bibinfo{pages}{187601} (\bibinfo{year}{2003}).

\bibitem[{\citenamefont{Anderson et~al.}(1979)\citenamefont{Anderson, Abrahams,
  and Ramakrishnan}}]{Anderson:1979}
\bibinfo{author}{\bibfnamefont{P.}~\bibnamefont{Anderson}},
  \bibinfo{author}{\bibfnamefont{E.}~\bibnamefont{Abrahams}}, \bibnamefont{and}
  \bibinfo{author}{\bibfnamefont{T.}~\bibnamefont{Ramakrishnan}},
  \bibinfo{journal}{Phys. Rev. Lett.} \textbf{\bibinfo{volume}{43}},
  \bibinfo{pages}{718} (\bibinfo{year}{1979}).

\bibitem[{\citenamefont{Gorkov et~al.}(1979)\citenamefont{Gorkov, Larkin, and
  Khmelnitskii}}]{Gorkov:1979}
\bibinfo{author}{\bibfnamefont{L.~P.} \bibnamefont{Gorkov}},
  \bibinfo{author}{\bibfnamefont{A.}~\bibnamefont{Larkin}}, \bibnamefont{and}
  \bibinfo{author}{\bibfnamefont{D.}~\bibnamefont{Khmelnitskii}},
  \bibinfo{journal}{Sov. Phys. JETP Lett.} \textbf{\bibinfo{volume}{30}},
  \bibinfo{pages}{228} (\bibinfo{year}{1979}).

\bibitem[{\citenamefont{Altshuler}(1985)}]{Altshuler:1985}
\bibinfo{author}{\bibfnamefont{B.}~\bibnamefont{Altshuler}},
  \bibinfo{journal}{Sov. Phys. JETP Lett.} \textbf{\bibinfo{volume}{41}},
  \bibinfo{pages}{648} (\bibinfo{year}{1985}).

\bibitem[{\citenamefont{Lee and Stone}(1985)}]{Lee:1985}
\bibinfo{author}{\bibfnamefont{P.}~\bibnamefont{Lee}} \bibnamefont{and}
  \bibinfo{author}{\bibfnamefont{A.}~\bibnamefont{Stone}},
  \bibinfo{journal}{Phys. Rev. Lett.} \textbf{\bibinfo{volume}{55}},
  \bibinfo{pages}{1622} (\bibinfo{year}{1985}).

\bibitem[{\citenamefont{Joos et~al.}(2003)\citenamefont{Joos, Zeh, Kieffer,
  Giulini, Kupsch, and Stamatescu}}]{Joos:book}
\bibinfo{author}{\bibfnamefont{E.}~\bibnamefont{Joos}},
  \bibinfo{author}{\bibfnamefont{H.}~\bibnamefont{Zeh}},
  \bibinfo{author}{\bibfnamefont{C.}~\bibnamefont{Kieffer}},
  \bibinfo{author}{\bibfnamefont{D.}~\bibnamefont{Giulini}},
  \bibinfo{author}{\bibfnamefont{J.}~\bibnamefont{Kupsch}}, \bibnamefont{and}
  \bibinfo{author}{\bibfnamefont{I.-O.} \bibnamefont{Stamatescu}},
  \emph{\bibinfo{title}{Decoherence and the appearance of a classical world in
  quantum theory}} (\bibinfo{publisher}{Springer}, \bibinfo{year}{2003}).

\bibitem[{\citenamefont{Xia et~al.}(2006)\citenamefont{Xia, Zwierzycki,
  Talanana, Kelly, and Bauer}}]{Xia:2006}
\bibinfo{author}{\bibfnamefont{K.}~\bibnamefont{Xia}},
  \bibinfo{author}{\bibfnamefont{M.}~\bibnamefont{Zwierzycki}},
  \bibinfo{author}{\bibfnamefont{M.}~\bibnamefont{Talanana}},
  \bibinfo{author}{\bibfnamefont{P.~J.} \bibnamefont{Kelly}}, \bibnamefont{and}
  \bibinfo{author}{\bibfnamefont{G.~E.~W.} \bibnamefont{Bauer}},
  \bibinfo{journal}{Phys. Rev. B} \textbf{\bibinfo{volume}{73}},
  \bibinfo{pages}{064420} (\bibinfo{year}{2006}).

\bibitem[{\citenamefont{Stiles and
  Zangwill}(2002{\natexlab{b}})}]{Stiles:2002b}
\bibinfo{author}{\bibfnamefont{M.}~\bibnamefont{Stiles}} \bibnamefont{and}
  \bibinfo{author}{\bibfnamefont{A.}~\bibnamefont{Zangwill}},
  \bibinfo{journal}{Phys. Rev. B} \textbf{\bibinfo{volume}{66}},
  \bibinfo{pages}{14407} (\bibinfo{year}{2002}{\natexlab{b}}).

\bibitem[{\citenamefont{Taniguchi
  et~al.}(2008{\natexlab{a}})\citenamefont{Taniguchi, Yakata, Imamura, and
  Ando}}]{Taniguchi:2008a}
\bibinfo{author}{\bibfnamefont{T.}~\bibnamefont{Taniguchi}},
  \bibinfo{author}{\bibfnamefont{S.}~\bibnamefont{Yakata}},
  \bibinfo{author}{\bibfnamefont{H.}~\bibnamefont{Imamura}}, \bibnamefont{and}
  \bibinfo{author}{\bibfnamefont{Y.}~\bibnamefont{Ando}},
  \bibinfo{journal}{Appl. Phys. Exp.} \textbf{\bibinfo{volume}{1}},
  \bibinfo{pages}{031302} (\bibinfo{year}{2008}{\natexlab{a}}).

\bibitem[{\citenamefont{Taniguchi
  et~al.}(2008{\natexlab{b}})\citenamefont{Taniguchi, Yakata, Imamura, and
  Ando}}]{Taniguchi:2008b}
\bibinfo{author}{\bibfnamefont{T.}~\bibnamefont{Taniguchi}},
  \bibinfo{author}{\bibfnamefont{S.}~\bibnamefont{Yakata}},
  \bibinfo{author}{\bibfnamefont{H.}~\bibnamefont{Imamura}}, \bibnamefont{and}
  \bibinfo{author}{\bibfnamefont{Y.}~\bibnamefont{Ando}}, \bibinfo{journal}{Mag
  IEEE} \textbf{\bibinfo{volume}{44}}, \bibinfo{pages}{2636}
  (\bibinfo{year}{2008}{\natexlab{b}}).

\bibitem[{\citenamefont{Taniguchi and Imamura}(2008)}]{Taniguchi:2008c}
\bibinfo{author}{\bibfnamefont{T.}~\bibnamefont{Taniguchi}} \bibnamefont{and}
  \bibinfo{author}{\bibfnamefont{H.}~\bibnamefont{Imamura}},
  \bibinfo{journal}{Phys. Rev. B} \textbf{\bibinfo{volume}{78}},
  \bibinfo{pages}{224421} (\bibinfo{year}{2008}).

\bibitem[{\citenamefont{Yang et~al.}(1995)\citenamefont{Yang, Holody, Loloee,
  Henry, Pratt, Jr, Schroeder, and Bass}}]{Yang:1995}
\bibinfo{author}{\bibfnamefont{Q.}~\bibnamefont{Yang}},
  \bibinfo{author}{\bibfnamefont{P.}~\bibnamefont{Holody}},
  \bibinfo{author}{\bibfnamefont{R.}~\bibnamefont{Loloee}},
  \bibinfo{author}{\bibfnamefont{L.~L.} \bibnamefont{Henry}},
  \bibinfo{author}{\bibfnamefont{W.~P.} \bibnamefont{Pratt}},
  \bibinfo{author}{\bibnamefont{Jr}}, \bibinfo{author}{\bibfnamefont{P.~A.}
  \bibnamefont{Schroeder}}, \bibnamefont{and}
  \bibinfo{author}{\bibfnamefont{J.}~\bibnamefont{Bass}},
  \bibinfo{journal}{Phys. Rev. B} \textbf{\bibinfo{volume}{51}},
  \bibinfo{pages}{3226} (\bibinfo{year}{1995}).

\bibitem[{\citenamefont{Xia et~al.}(2001)\citenamefont{Xia, Kelly, Bauer,
  Turek, Kudrnovsk{\'y}, and Drchal}}]{Xia:2001}
\bibinfo{author}{\bibfnamefont{K.}~\bibnamefont{Xia}},
  \bibinfo{author}{\bibfnamefont{P.~J.} \bibnamefont{Kelly}},
  \bibinfo{author}{\bibfnamefont{G.~E.~W.} \bibnamefont{Bauer}},
  \bibinfo{author}{\bibfnamefont{I.}~\bibnamefont{Turek}},
  \bibinfo{author}{\bibfnamefont{J.}~\bibnamefont{Kudrnovsk{\'y}}},
  \bibnamefont{and} \bibinfo{author}{\bibfnamefont{V.}~\bibnamefont{Drchal}},
  \bibinfo{journal}{Phys. Rev. B} \textbf{\bibinfo{volume}{63}},
  \bibinfo{pages}{064407} (\bibinfo{year}{2001}).

\bibitem[{\citenamefont{Park et~al.}(2000)\citenamefont{Park, Baxter, Steenwyk,
  Moraru, Pratt, and Bass}}]{Park:2000}
\bibinfo{author}{\bibfnamefont{W.}~\bibnamefont{Park}},
  \bibinfo{author}{\bibfnamefont{D.~V.} \bibnamefont{Baxter}},
  \bibinfo{author}{\bibfnamefont{S.}~\bibnamefont{Steenwyk}},
  \bibinfo{author}{\bibfnamefont{I.}~\bibnamefont{Moraru}},
  \bibinfo{author}{\bibfnamefont{W.~P.} \bibnamefont{Pratt}}, \bibnamefont{and}
  \bibinfo{author}{\bibfnamefont{J.}~\bibnamefont{Bass}},
  \bibinfo{journal}{Phys. Rev. B} \textbf{\bibinfo{volume}{62}},
  \bibinfo{pages}{1178} (\bibinfo{year}{2000}).

\bibitem[{\citenamefont{van Wees et~al.}(1988)\citenamefont{van Wees, van
  Houten, Beenakker, Williamson, Kouwenhoven, van~der Marel, and
  Foxon}}]{Wees:1988}
\bibinfo{author}{\bibfnamefont{B.~J.} \bibnamefont{van Wees}},
  \bibinfo{author}{\bibfnamefont{H.}~\bibnamefont{van Houten}},
  \bibinfo{author}{\bibfnamefont{C.~W.~J.} \bibnamefont{Beenakker}},
  \bibinfo{author}{\bibfnamefont{J.~G.} \bibnamefont{Williamson}},
  \bibinfo{author}{\bibfnamefont{L.~P.} \bibnamefont{Kouwenhoven}},
  \bibinfo{author}{\bibfnamefont{D.}~\bibnamefont{van~der Marel}},
  \bibnamefont{and} \bibinfo{author}{\bibfnamefont{C.~T.} \bibnamefont{Foxon}},
  \bibinfo{journal}{Phys. Rev. Lett.} \textbf{\bibinfo{volume}{60}},
  \bibinfo{pages}{848} (\bibinfo{year}{1988}).

\bibitem[{\citenamefont{Wharam et~al.}(1988)\citenamefont{Wharam, Thornton,
  Newbury, Pepper, Ahmed, Frost, Hasko, Peacock, Ritchie, and
  Jones}}]{Wharam:1988}
\bibinfo{author}{\bibfnamefont{D.}~\bibnamefont{Wharam}},
  \bibinfo{author}{\bibfnamefont{T.}~\bibnamefont{Thornton}},
  \bibinfo{author}{\bibfnamefont{R.}~\bibnamefont{Newbury}},
  \bibinfo{author}{\bibfnamefont{M.}~\bibnamefont{Pepper}},
  \bibinfo{author}{\bibfnamefont{H.}~\bibnamefont{Ahmed}},
  \bibinfo{author}{\bibfnamefont{J.}~\bibnamefont{Frost}},
  \bibinfo{author}{\bibfnamefont{D.}~\bibnamefont{Hasko}},
  \bibinfo{author}{\bibfnamefont{D.}~\bibnamefont{Peacock}},
  \bibinfo{author}{\bibfnamefont{D.}~\bibnamefont{Ritchie}}, \bibnamefont{and}
  \bibinfo{author}{\bibfnamefont{G.}~\bibnamefont{Jones}}, \bibinfo{journal}{J.
  Phys. C} \textbf{\bibinfo{volume}{21}}, \bibinfo{pages}{L209}
  (\bibinfo{year}{1988}).

\bibitem[{\citenamefont{Sanvito et~al.}(1999)\citenamefont{Sanvito, Lambert,
  Jefferson, and Bratkovsky}}]{Sanvito:1999}
\bibinfo{author}{\bibfnamefont{S.}~\bibnamefont{Sanvito}},
  \bibinfo{author}{\bibfnamefont{C.}~\bibnamefont{Lambert}},
  \bibinfo{author}{\bibfnamefont{J.}~\bibnamefont{Jefferson}},
  \bibnamefont{and}
  \bibinfo{author}{\bibfnamefont{A.}~\bibnamefont{Bratkovsky}},
  \bibinfo{journal}{Phys. Rev. B} \textbf{\bibinfo{volume}{59}},
  \bibinfo{pages}{11936} (\bibinfo{year}{1999}).

\bibitem[{\citenamefont{Kazymyrenko and Waintal}(2008)}]{Kazymyrenko:2008}
\bibinfo{author}{\bibfnamefont{K.}~\bibnamefont{Kazymyrenko}} \bibnamefont{and}
  \bibinfo{author}{\bibfnamefont{X.}~\bibnamefont{Waintal}},
  \bibinfo{journal}{Phys. Rev. B} \textbf{\bibinfo{volume}{77}},
  \bibinfo{pages}{115119} (\bibinfo{year}{2008}).

\bibitem[{\citenamefont{Rychkova et~al.}()\citenamefont{Rychkova, Rychkov,
  Kazymyrenko, Borlenghi, and Waintal}}]{Rychkova:2010}
\bibinfo{author}{\bibfnamefont{I.}~\bibnamefont{Rychkova}},
  \bibinfo{author}{\bibfnamefont{V.}~\bibnamefont{Rychkov}},
  \bibinfo{author}{\bibfnamefont{K.}~\bibnamefont{Kazymyrenko}},
  \bibinfo{author}{\bibfnamefont{S.}~\bibnamefont{Borlenghi}},
  \bibnamefont{and} \bibinfo{author}{\bibfnamefont{X.}~\bibnamefont{Waintal}},
  \bibinfo{note}{arXiv:1010.2627}.

\bibitem[{\citenamefont{Caroli et~al.}(1971)\citenamefont{Caroli, Combescot,
  Nozi\`eres, and Saint-James}}]{Caroli:1971}
\bibinfo{author}{\bibfnamefont{C.}~\bibnamefont{Caroli}},
  \bibinfo{author}{\bibfnamefont{R.}~\bibnamefont{Combescot}},
  \bibinfo{author}{\bibfnamefont{P.}~\bibnamefont{Nozi\`eres}},
  \bibnamefont{and}
  \bibinfo{author}{\bibfnamefont{D.}~\bibnamefont{Saint-James}},
  \bibinfo{journal}{J. Phys. C} \textbf{\bibinfo{volume}{4}},
  \bibinfo{pages}{916} (\bibinfo{year}{1971}).

\bibitem[{\citenamefont{Meir and Wingreen}(1992)}]{Meir:1992}
\bibinfo{author}{\bibfnamefont{Y.}~\bibnamefont{Meir}} \bibnamefont{and}
  \bibinfo{author}{\bibfnamefont{N.}~\bibnamefont{Wingreen}},
  \bibinfo{journal}{Phys. Rev. Lett.} \textbf{\bibinfo{volume}{68}},
  \bibinfo{pages}{2512} (\bibinfo{year}{1992}).

\bibitem[{\citenamefont{Fisher and Lee}(1981)}]{Fisher:1981}
\bibinfo{author}{\bibfnamefont{D.}~\bibnamefont{Fisher}} \bibnamefont{and}
  \bibinfo{author}{\bibfnamefont{P.}~\bibnamefont{Lee}},
  \bibinfo{journal}{Phys. Rev. B} \textbf{\bibinfo{volume}{23}},
  \bibinfo{pages}{6851} (\bibinfo{year}{1981}).

\bibitem[{\citenamefont{Stiles}(1996{\natexlab{a}})}]{Stiles:1996a}
\bibinfo{author}{\bibfnamefont{M.~D.} \bibnamefont{Stiles}},
  \bibinfo{journal}{Phys. Rev. B} \textbf{\bibinfo{volume}{54}},
  \bibinfo{pages}{14679} (\bibinfo{year}{1996}{\natexlab{a}}).

\bibitem[{\citenamefont{Stiles}(1996{\natexlab{b}})}]{Stiles:1996b}
\bibinfo{author}{\bibfnamefont{M.~D.} \bibnamefont{Stiles}},
  \bibinfo{journal}{J. Appl. Phys.} \textbf{\bibinfo{volume}{79}},
  \bibinfo{pages}{5805} (\bibinfo{year}{1996}{\natexlab{b}}).

\bibitem[{\citenamefont{Berger}(1996)}]{Berger:1996}
\bibinfo{author}{\bibfnamefont{L.}~\bibnamefont{Berger}},
  \bibinfo{journal}{Phys. Rev. B} \textbf{\bibinfo{volume}{54}},
  \bibinfo{pages}{9353} (\bibinfo{year}{1996}).

\bibitem[{\citenamefont{Slonczewski}(1996)}]{Slonczewski:1996}
\bibinfo{author}{\bibfnamefont{J.~C.} \bibnamefont{Slonczewski}},
  \bibinfo{journal}{JMMM} \textbf{\bibinfo{volume}{62}}, \bibinfo{pages}{L1}
  (\bibinfo{year}{1996}).

\bibitem[{\citenamefont{Manschot et~al.}(2004)\citenamefont{Manschot, Brataas,
  and Bauer}}]{Manschot:2004}
\bibinfo{author}{\bibfnamefont{J.}~\bibnamefont{Manschot}},
  \bibinfo{author}{\bibfnamefont{A.}~\bibnamefont{Brataas}}, \bibnamefont{and}
  \bibinfo{author}{\bibfnamefont{G.}~\bibnamefont{Bauer}},
  \bibinfo{journal}{Phys. Rev. B} \textbf{\bibinfo{volume}{69}},
  \bibinfo{pages}{092407} (\bibinfo{year}{2004}).

\bibitem[{\citenamefont{Barna\'{s} et~al.}(2005)\citenamefont{Barna\'{s}, Fert,
  Gmitra, Weymann, and Dugaev}}]{Barnas:2005}
\bibinfo{author}{\bibfnamefont{J.}~\bibnamefont{Barna\'{s}}},
  \bibinfo{author}{\bibfnamefont{A.}~\bibnamefont{Fert}},
  \bibinfo{author}{\bibfnamefont{M.}~\bibnamefont{Gmitra}},
  \bibinfo{author}{\bibfnamefont{I.}~\bibnamefont{Weymann}}, \bibnamefont{and}
  \bibinfo{author}{\bibfnamefont{V.}~\bibnamefont{Dugaev}},
  \bibinfo{journal}{Phys. Rev. B} \textbf{\bibinfo{volume}{72}},
  \bibinfo{pages}{024426} (\bibinfo{year}{2005}).

\bibitem[{\citenamefont{Gmitra and Barnas}(2006)}]{Gmitra:2006}
\bibinfo{author}{\bibfnamefont{M.}~\bibnamefont{Gmitra}} \bibnamefont{and}
  \bibinfo{author}{\bibfnamefont{J.}~\bibnamefont{Barnas}},
  \bibinfo{journal}{Phys. Rev. Lett.} \textbf{\bibinfo{volume}{96}},
  \bibinfo{pages}{207205} (\bibinfo{year}{2006}).

\bibitem[{\citenamefont{Boulle et~al.}(2007)\citenamefont{Boulle, Cros,
  Grollier, Pereira, Deranlot, Petroff, Faini, Barna, and Fert}}]{Boulle:2007}
\bibinfo{author}{\bibfnamefont{O.}~\bibnamefont{Boulle}},
  \bibinfo{author}{\bibfnamefont{V.}~\bibnamefont{Cros}},
  \bibinfo{author}{\bibfnamefont{J.}~\bibnamefont{Grollier}},
  \bibinfo{author}{\bibfnamefont{L.~G.} \bibnamefont{Pereira}},
  \bibinfo{author}{\bibfnamefont{C.}~\bibnamefont{Deranlot}},
  \bibinfo{author}{\bibfnamefont{F.}~\bibnamefont{Petroff}},
  \bibinfo{author}{\bibfnamefont{G.}~\bibnamefont{Faini}},
  \bibinfo{author}{\bibfnamefont{J.}~\bibnamefont{Barna}}, \bibnamefont{and}
  \bibinfo{author}{\bibfnamefont{A.}~\bibnamefont{Fert}},
  \bibinfo{journal}{Nature Phys.} \textbf{\bibinfo{volume}{3}},
  \bibinfo{pages}{492} (\bibinfo{year}{2007}).

\bibitem[{\citenamefont{Gmitra and Barna\'{s}}(2009)}]{Gmitra:2009}
\bibinfo{author}{\bibfnamefont{M.}~\bibnamefont{Gmitra}} \bibnamefont{and}
  \bibinfo{author}{\bibfnamefont{J.}~\bibnamefont{Barna\'{s}}},
  \bibinfo{journal}{Phys. Rev. B} \textbf{\bibinfo{volume}{79}}
  (\bibinfo{year}{2009}).

\bibitem[{\citenamefont{Petitjean and Waintal}()}]{Petitjean:2011}
\bibinfo{author}{\bibfnamefont{C.}~\bibnamefont{Petitjean}} \bibnamefont{and}
  \bibinfo{author}{\bibfnamefont{X.}~\bibnamefont{Waintal}}, \bibinfo{note}{in
  preparation.}

\bibitem[{\citenamefont{Shurbet et~al.}(1974)\citenamefont{Shurbet, Lewis, and
  Boullion}}]{Shurbet:1974}
\bibinfo{author}{\bibfnamefont{G.}~\bibnamefont{Shurbet}},
  \bibinfo{author}{\bibfnamefont{T.}~\bibnamefont{Lewis}}, \bibnamefont{and}
  \bibinfo{author}{\bibfnamefont{T.}~\bibnamefont{Boullion}},
  \bibinfo{journal}{The Ohio Journal of Science} \textbf{\bibinfo{volume}{74}},
  \bibinfo{pages}{273} (\bibinfo{year}{1974}).

\end{thebibliography}
\end{document}